\def\year{2020}\relax

\documentclass[letterpaper]{article} 
\usepackage{aaai20}  
\usepackage{times}  
\usepackage{helvet} 
\usepackage{courier}  
\usepackage[hyphens]{url}  
\usepackage{graphicx} 
\usepackage{graphicx}  
\usepackage{xcolor}
\usepackage{listings}
\usepackage{microtype}
\usepackage{subcaption}
\usepackage{booktabs} 

\usepackage{amsmath,amsfonts,bm}

\def\eqref#1{equation~\ref{#1}}

\def\1{\bm{1}}

\def\ra{{\textnormal{a}}}
\def\rb{{\textnormal{b}}}

\def\rd{{\textnormal{d}}}

\def\rp{{\textnormal{p}}}
\def\rq{{\textnormal{q}}}

\def\rt{{\textnormal{t}}}

\def\rvs{{\mathbf{s}}}

\def\rvx{{\mathbf{x}}}
\def\rvy{{\mathbf{y}}}

\def\vs{{\bm{s}}}
\def\vt{{\bm{t}}}

\def\vx{{\bm{x}}}

\DeclareMathAlphabet{\mathsfit}{\encodingdefault}{\sfdefault}{m}{sl}
\SetMathAlphabet{\mathsfit}{bold}{\encodingdefault}{\sfdefault}{bx}{n}

\begin{document}
\title{Generating Realistic Stock Market Order Streams}
\author{Junyi Li,\textsuperscript{\rm 1} Xintong Wang,\textsuperscript{\rm 2} Yaoyang Lin,\textsuperscript{\rm 3} Arunesh Sinha,\textsuperscript{\rm 4} Michael P. Wellman\textsuperscript{\rm 2} \\
\textsuperscript{\rm 1}{University of Pittsburgh}, \textsuperscript{\rm 2}{University of Michigan}, \\
\textsuperscript{\rm 3}{Harvard University}, 
\textsuperscript{\rm 4}{Singapore Management University} \\
jul116@pitt.edu, xintongw@umich.edu, yaoyanglin@g.harvard.edu,  aruneshs@smu.edu.sg, wellman@umich.edu
}

\maketitle
\begin{abstract}
We propose an approach to generate realistic and high-fidelity stock market data based on generative adversarial networks (GANs).
Our Stock-GAN model employs a conditional Wasserstein GAN to capture history dependence of orders. 
The generator design includes specially crafted aspects including components that approximate the market's auction mechanism, augmenting the order history with order-book constructions to improve the generation task. We perform an ablation study to verify the usefulness of aspects of our network structure.
We provide a mathematical characterization of distribution learned by the generator. We also propose statistics to measure the quality of generated orders.
We test our approach with synthetic and actual market data, compare to many baseline generative models, and find the generated data to be close to real data. 
\end{abstract}

\section{Introduction}
Financial markets are among the most well-studied and closely watched complex multiagent systems in existence. 
Well-functioning financial markets are critical to the operation of a complex global economy, and small changes in the efficiency or stability of such markets can have enormous ramifications. 
Accurate modeling of financial markets can support improved design and regulation of these critical institutions.
There is a vast literature on financial market modeling, though still a large gap between the state-of-art and the ideal.
Analytic approaches provide insight through highly stylized model forms.
Agent-based models accommodate greater dynamic complexity, and are often able to reproduce ``stylized facts'' of real-world markets \cite{LeBaron06}.
Currently lacking, however, is a simulation capable of producing market data at high fidelity and high realism.
Our aim is to develop such a model, to support a range of market design and analysis problems.
This work provides a first step, learning a high-fidelity generator from real stock market data streams.

Our \emph{main contribution} is Stock-GAN: an approach to produce realistic stock market order streams from  real market data.
We utilize a conditional Wasserstein GAN (WGAN) \cite{arjovsky2017wasserstein,mirza2014conditional} to capture the time-dependence of order streams, with both the generator and critic conditional on history of orders. 
The \emph{main innovation} in the Stock-GAN network architecture lies in two deliberately crafted features of the generator. 
The first is a separate neural network that is used to approximate the double auction mechanism underlying stock exchanges. 
This pre-trained network is embedded in the generator enabling it to model order processing and transaction generation. The second feature is the inclusion of order-book information in the conditioning history of the network. 
The order book captures the key features of market state that are not directly apparent from order history segments.

Our second contribution is a mathematical characterization of the distribution learned by the generator. We show that our designed generator models the stock market data stream as arising from a stochastic process with finite memory dependence. The stochastic process view also makes precise the conditional distribution that the generator is learning as well the joint distribution that the critic of the WGAN distinguishes by estimating the earth mover's distance. The stochastic process has no closed form representation, which necessitates the use of a neural network to learn it.

Finally, we experiment with synthetic and real market data. The synthetic data is produced using a stock market simulator that has been used in several agent-based financial studies~\cite{wellman2017strategic}, but is far from real market data.
The real market data was obtained from OneMarketData, a financial data provider.
We \emph{propose} five statistics for evaluating stock market data, such as the distribution of price and quantity of orders, inter-arrival times of orders, and the best bid and best ask evolution over time. We compare against other baseline generative models such as \emph{recurrent conditional} variational auto-encoder (VAE) and DCGAN instead of WGAN within Stock-GAN. We perform an ablation study showing the usefulness of our generator structure design as elaborated above. Overall, Stock-GAN is able to best generate realistic data compared to the alternatives. An appendix in the full version provides all additional results and code for our work.

\begin{figure*}[t]
\centering
\includegraphics[width=0.79\linewidth]{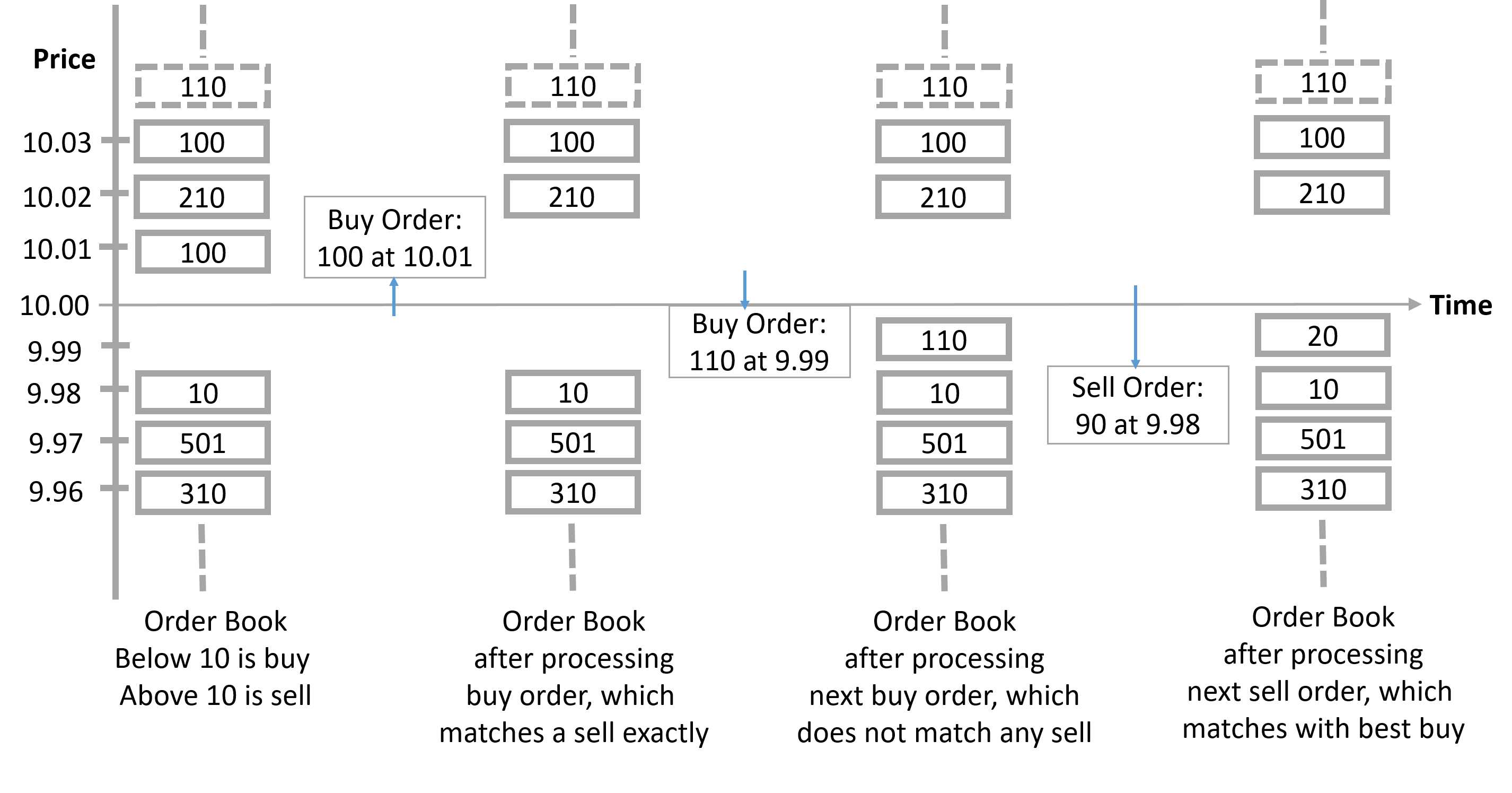}
\caption{Representation and evolution of a limit order book.}\label{CDAFig}
\end{figure*}

\section{Related Work and Background}
WGAN is a well-known GAN variant \cite{goodfellow2014generative,arjovsky2017wasserstein}. Most prior work on generation of sequences using GANs has been in the domain of text generation \cite{press2017language,zhang2017adversarial}. 
However, since the space of word representations is not continuous, the semantics change with nearby word representation, and given a lack of agreement on the metrics for measuring goodness of sentences, producing good quality text using GANs is still an active area of research. 
Stock market data does not suffer from this representation problem but the history dependence for stock markets can be much longer than for text generation. There are many advanced proposals to deal with long term dependence~\cite{neil2016phased,chang2017dilated,yu2017long}, however, we find that our use of LSTMs with conditional WGAN performs quite good with little tuning of hyperparameters.
Xiao et al.\ \shortcite{xiao2017wasserstein,Xiao2018LearningCG} introduced GAN-based methods for generating point processes; they generate the time for transaction events in stock markets. Other work aim to generate transaction prices in a stock market~\cite{da2019towards,koshiyama2019generative,zhang2019stock,wiese2019quant}. Our problem is richer and harder as we aim to generate the actual limit orders including time, order type, price, and quantity information.

Deep neural networks and machine learning techniques have been used on financial data mostly for prediction of transaction price \cite{M20181351,bao2017deep,qian2017financial,zhang2017stock} and for prediction of actual returns \cite{abe2018deep}. 
As stated, our goal is not market prediction per se, but rather market modeling.
Whereas the problems of learning to predict and generate may overlap (e.g., both aim to capture regularity in the domain), the evaluation criteria and end product are quite distinct. GANs have been used for generation of customer buy orders in e-commerce setting~\cite{shi2018virtual,kumar2018ecommercegan}, however, stock market orders are much more complex with buys, sells, and cancellations; further we attempt to ensure realism of higher level dynamics like the best bid and ask evolution over time.

\begin{figure*}[t]
\begin{subfigure}{0.14\linewidth}
\includegraphics[width=\linewidth]{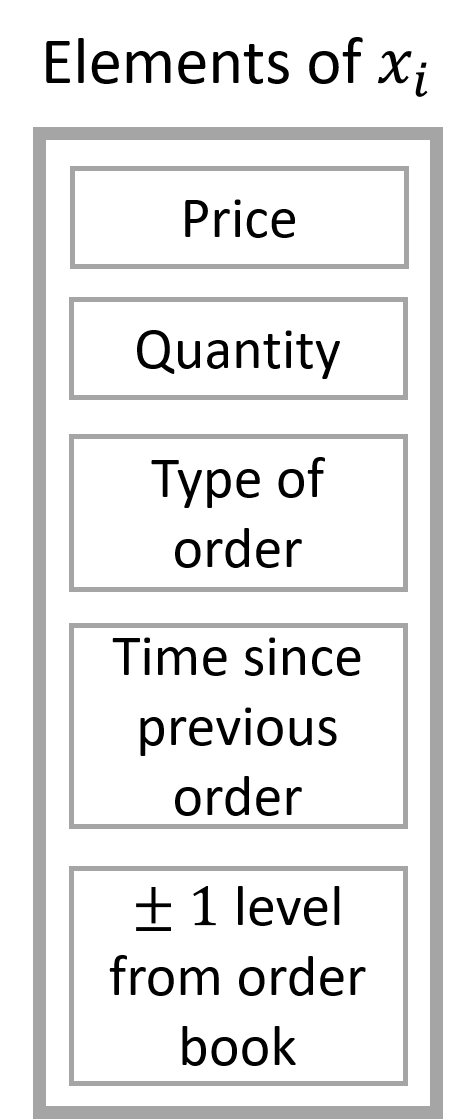}
\caption{$\rvx_i$}\label{1a}\hfill
\end{subfigure}
\begin{subfigure}{.42\linewidth}
\includegraphics[width=\linewidth]{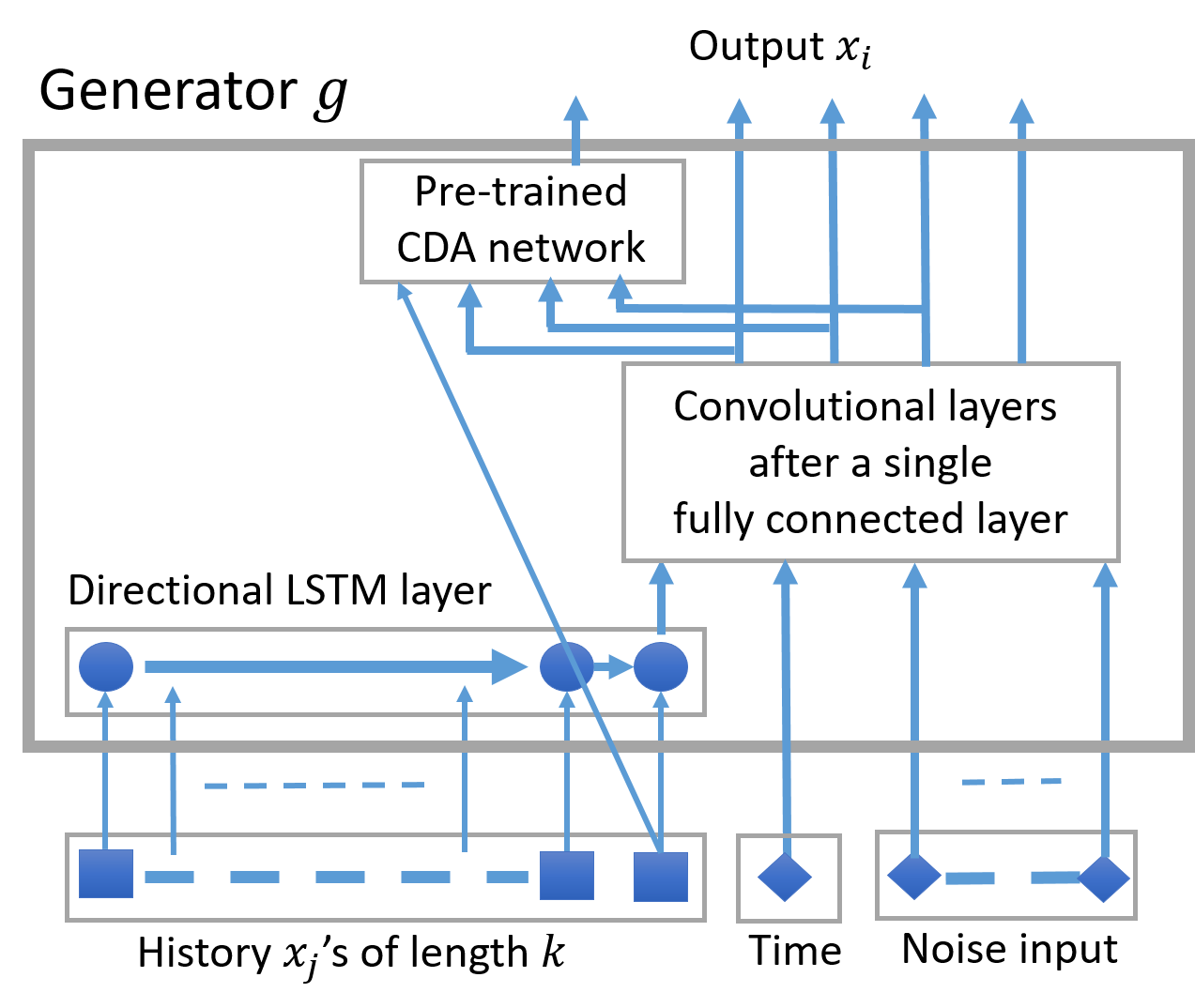}
\caption{Generator}\hfill
\end{subfigure}
\begin{subfigure}{.42\linewidth}
\includegraphics[width=\linewidth]{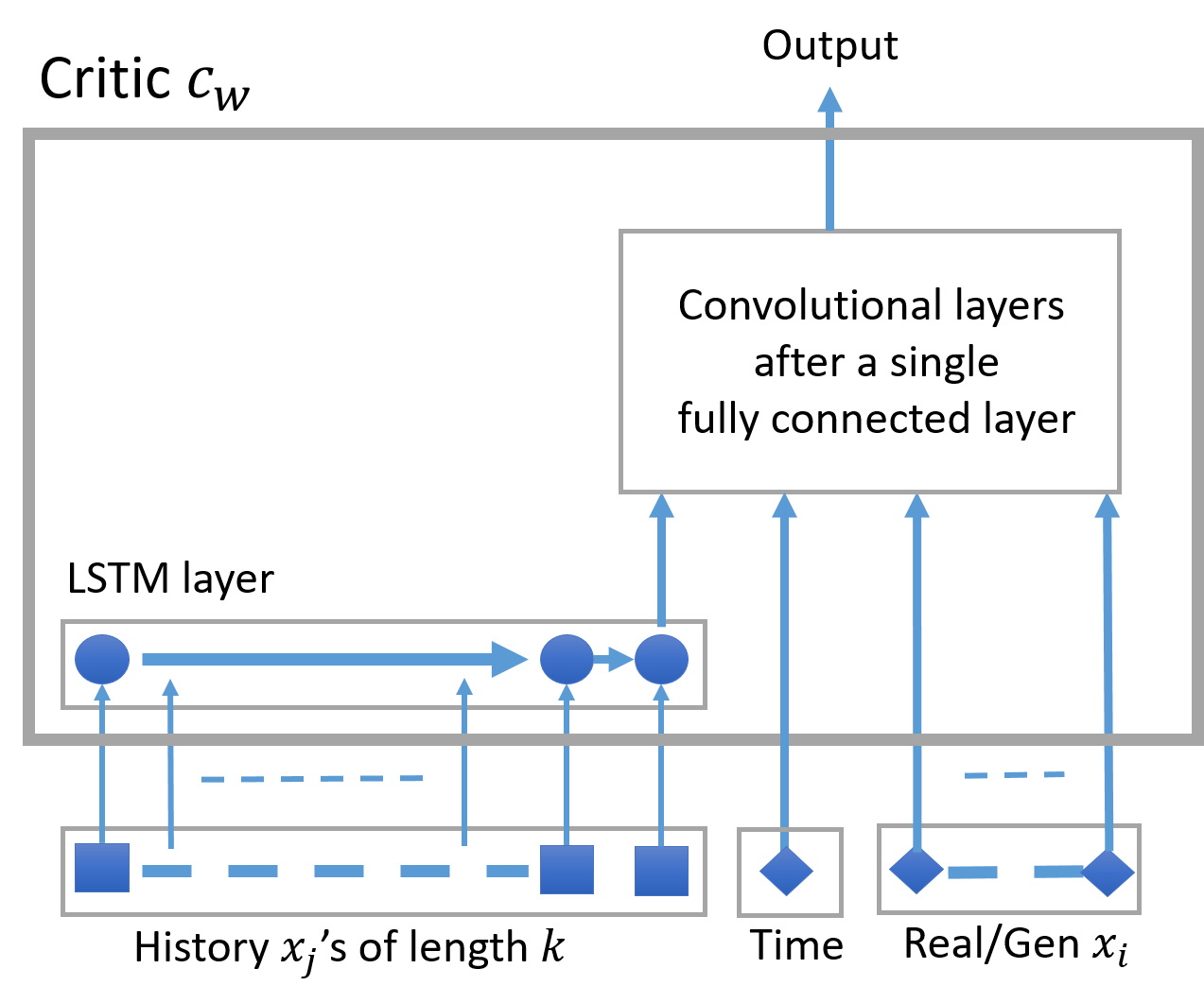}
\caption{Critic}\hfill
\end{subfigure}
\caption{Stock-GAN architecture}
\label{GAN}
\hfill
\end{figure*}

\textbf{Limit order books} 
The stock market is a venue where equities or stocks of publicly held companies are traded.
Nearly all stock markets follow the \emph{continuous double auction} (CDA) mechanism \cite{freidman1993double}.  
Traders submit bids, or \textit{limit orders}, specifying
the maximum price at which they would be
willing to buy a specified quantity of a stock, or the minimum price at which they would be willing to sell a quantity. %
\footnote{Hence, the CDA is often referred to as a limit-order market in the finance literature \cite{Abergel16}.} 
The \textit{order book} is a store that maintains the set of active orders: those submitted but not yet transacted or canceled. 
CDAs are continuous in the sense that when a new order matches an existing (incumbent) order in the order book, the market
clears immediately and the trade is executed at the price of the incumbent order---which is then removed from the order book. Orders may be submitted at any time, and a buy order matches and transacts with a sell order when their respective limits are mutually satisfied. 
For example, as shown in Figure~\ref{CDAFig}, if a buy order with price \$10.01 and quantity 100 arrives and the best sell offer in the order book has the same price and quantity, then they match exactly and transact.
As shown, the next buy order does not match any sell, and the following sell order partially matches what is then the best buy in the order book.

The limit order book maintains the current active orders in the market (or the state of the market), which can be described in terms of the quantity offered to buy or sell across the range of price levels. 
Each order arrival changes the market state, recorded as an update to the order book. 
After processing any arrived order every buy price level is higher than all sell price levels, and the \emph{best bid} refers to the lowest buy price level  and the \emph{best ask} refers to the highest sell price level. See Figure~\ref{CDAFig} for an illustration. The order book is often approximated by few (e.g., ten) price levels above the best bid and ten price levels below the best ask; as these prices are typically the ones that dictate the transactions in the market.
There are various kinds of traders in a stock market, ranging from individual investors to large investing firms. Thus, there is a wide variation in the nature of orders submitted.
We aim to generate streams of orders that are close in aggregate (not per trader) to real order streams for a given stock. 
We focus on generating orders and not transactions, as the CDA mechanism is deterministic and transactions can be determined exactly given a stream of orders. In fact, we model the CDA as a fixed (and separately learned) neural network within the generation process. In this work, we limit ourselves to limit orders as we do not have access to richer order types such as iceberg or bracket orders.

\section{Stock-GAN}
We view the stock market orders for a given chunk of time of day $\Delta \rt$ as a collection of vector valued random variable $\{\rvx_i\}_{i\in N}$ indexed by the limit order sequence number in $N = \{1, \ldots, n\}$. $\{\rvx_i\}$ corresponds to the $i^{th}$ limit order, but, includes more information than the limit order such as the current best bid and best ask. 
The components of the random vector $\rvx_i$ include the time interval $\rd_i$, type of order $\rt_i$, limit order price $\rp_i$, limit order quantity $\rq_i$, and the best bid $\ra_i$ and best ask $\rb_i$. 
The time interval $\rd_i$ specifies the difference in time between the current order $i$ and previous order $i-1$ (in precision of milliseconds); the range of $\rd_i$ is finite. 
The type of order can be buy, sell, cancel buy, or cancel sell (represented in two bits). The price and quantity are restricted to lie within finite bounds.
The price range is discretized in units of US cents and the quantity range is discretized in units of the equity (non-negative integers).

\begin{figure*}[t!]
\begin{subfigure}{.3\linewidth}\includegraphics[width=\linewidth]{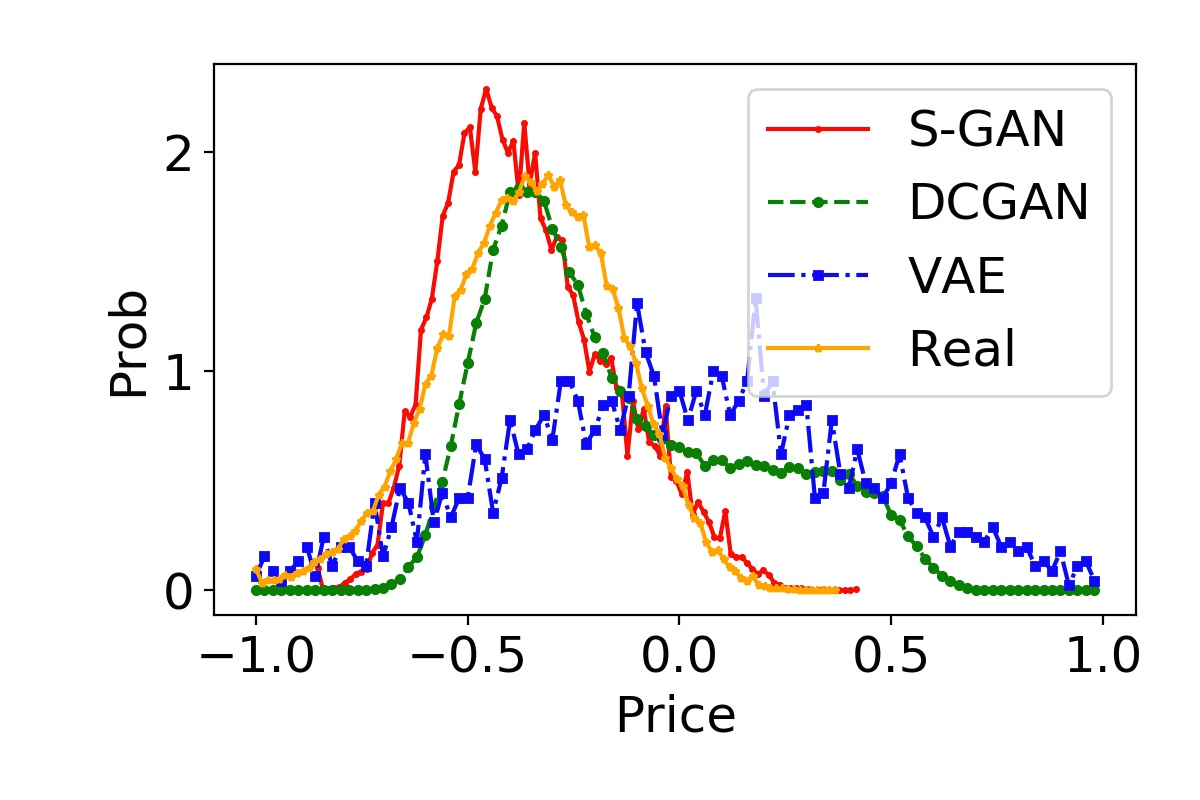}\caption{Synthetic price distribution}\label{fig:syn_price}\hfill\end{subfigure}
\begin{subfigure}{.3\linewidth}\includegraphics[width=\linewidth]{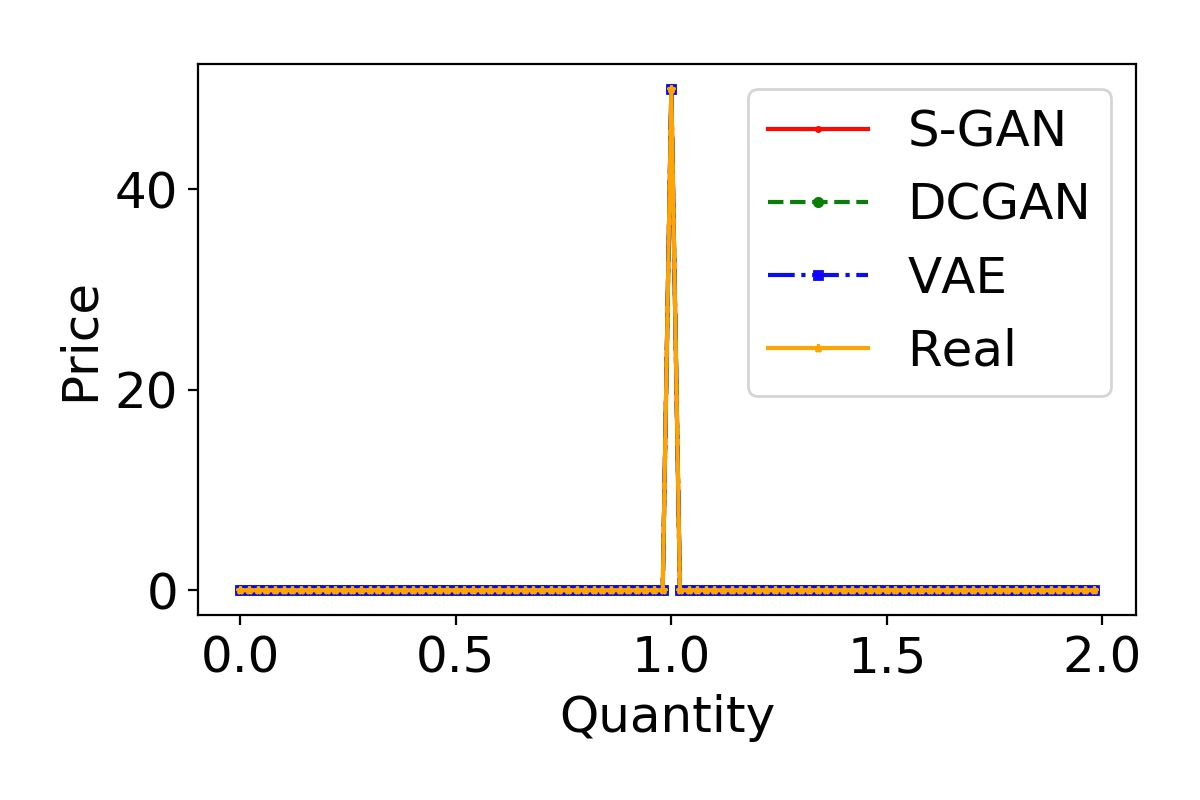}\caption{Synthetic quantity distribution}\label{fig:syn_qnt}\hfill\end{subfigure}
\begin{subfigure}{.3\linewidth}\includegraphics[width=\linewidth]{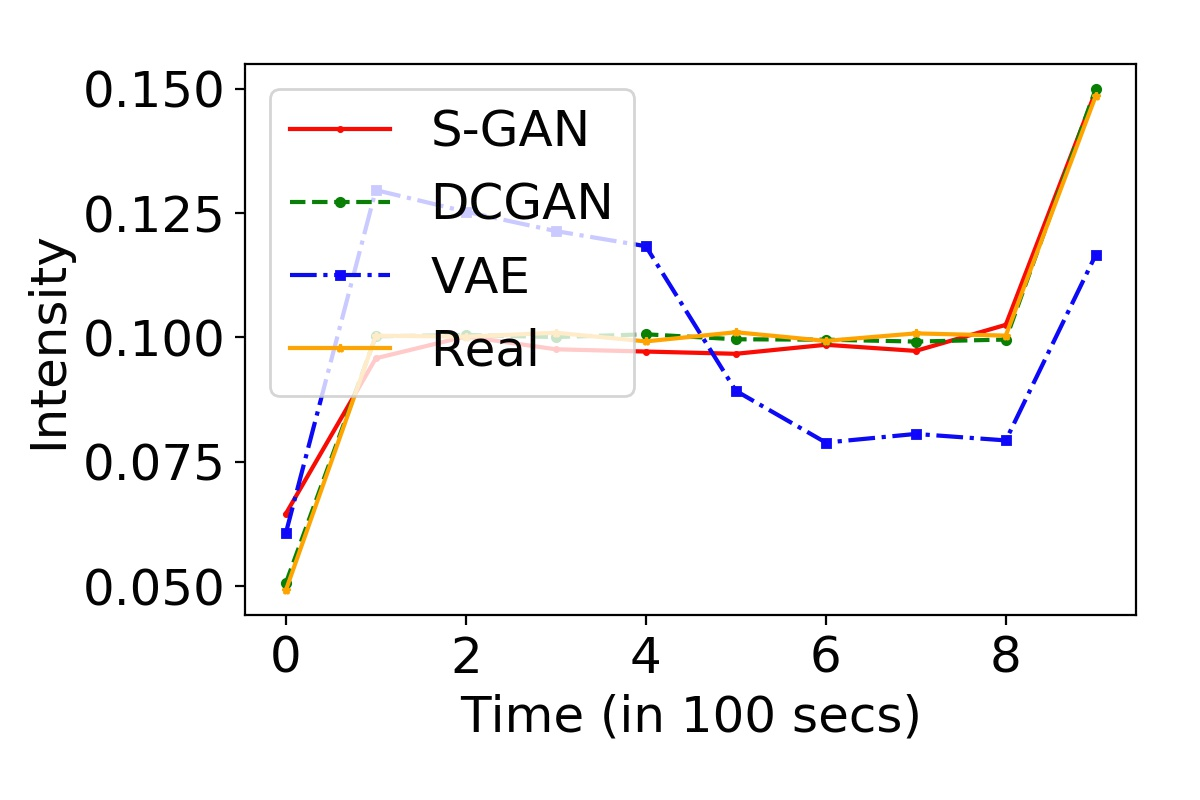}\caption{Synthetic intensity}\label{fig:syn_int}\hfill\end{subfigure}
\hspace*{\fill}
\hspace*{\fill}\\
\begin{subfigure}{.24\linewidth}\includegraphics[width=\linewidth]{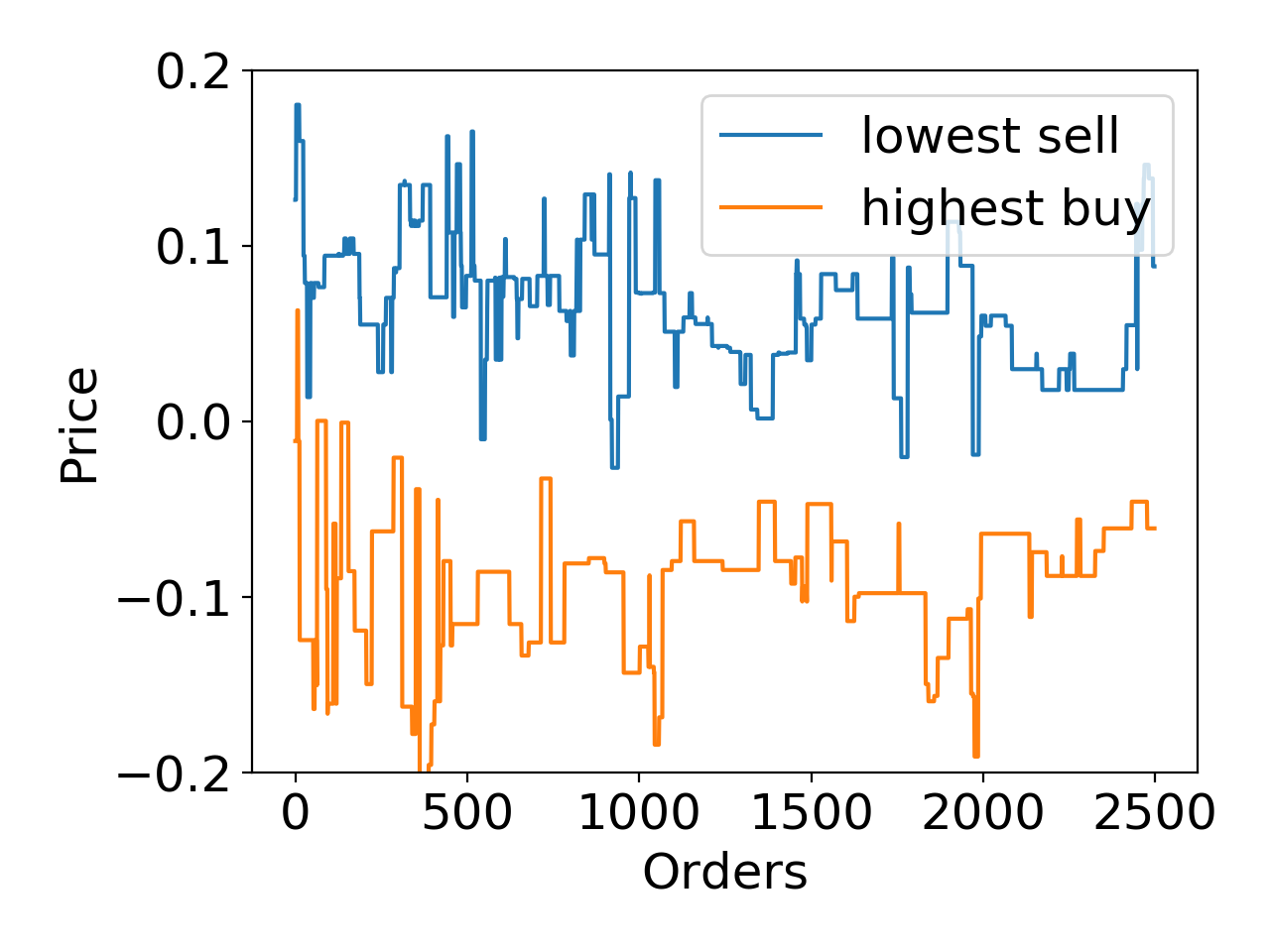}\caption{Synthetic best bid/ask}\label{fig:syn_bestba}\hfill\end{subfigure}
\begin{subfigure}{.24\linewidth}\includegraphics[width=\linewidth]{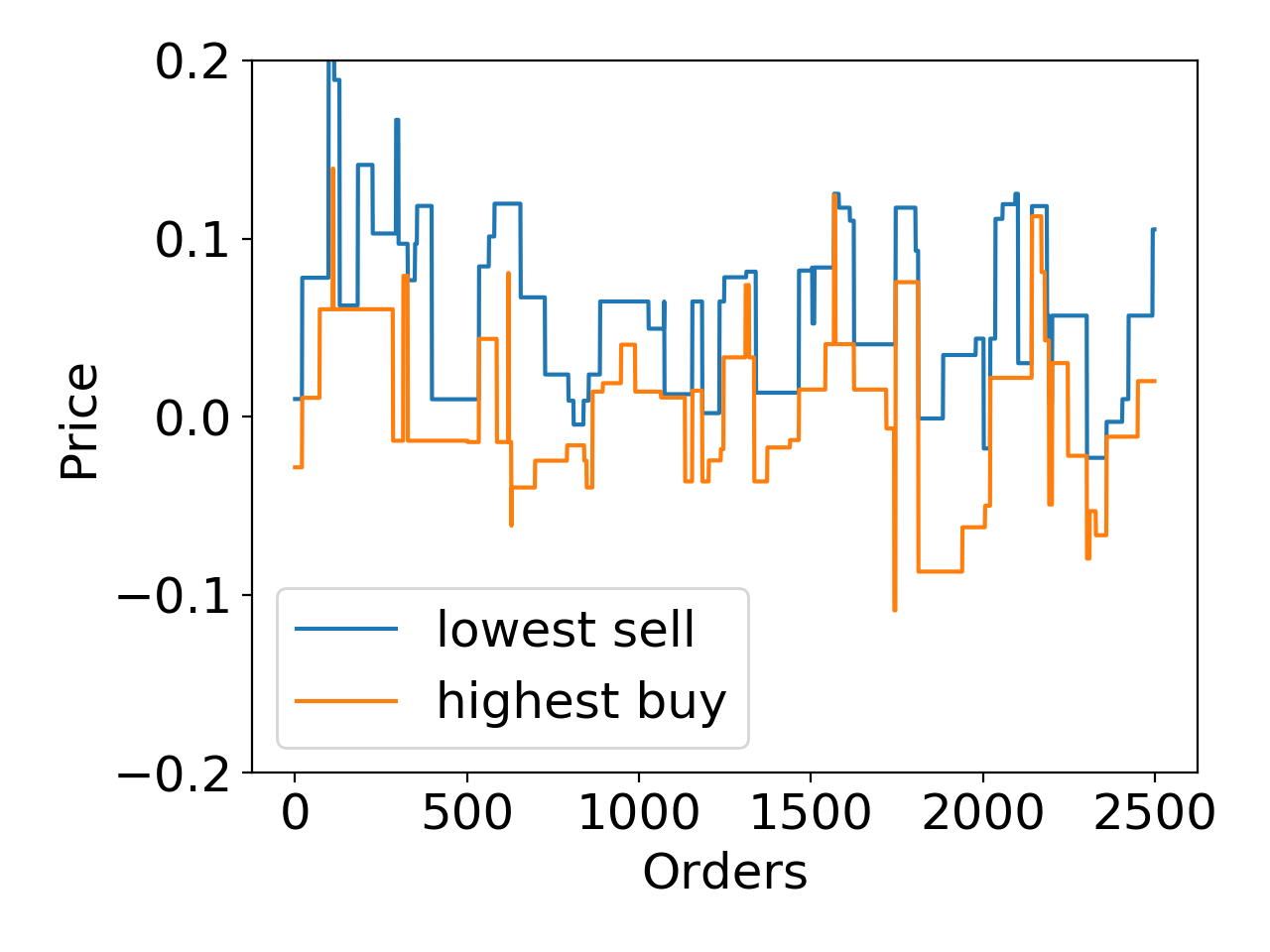}\caption{Stock-GAN best bid/ask}\label{fig:syn_bestbagen}\hfill\end{subfigure}
\begin{subfigure}{.24\linewidth}\includegraphics[width=\linewidth]{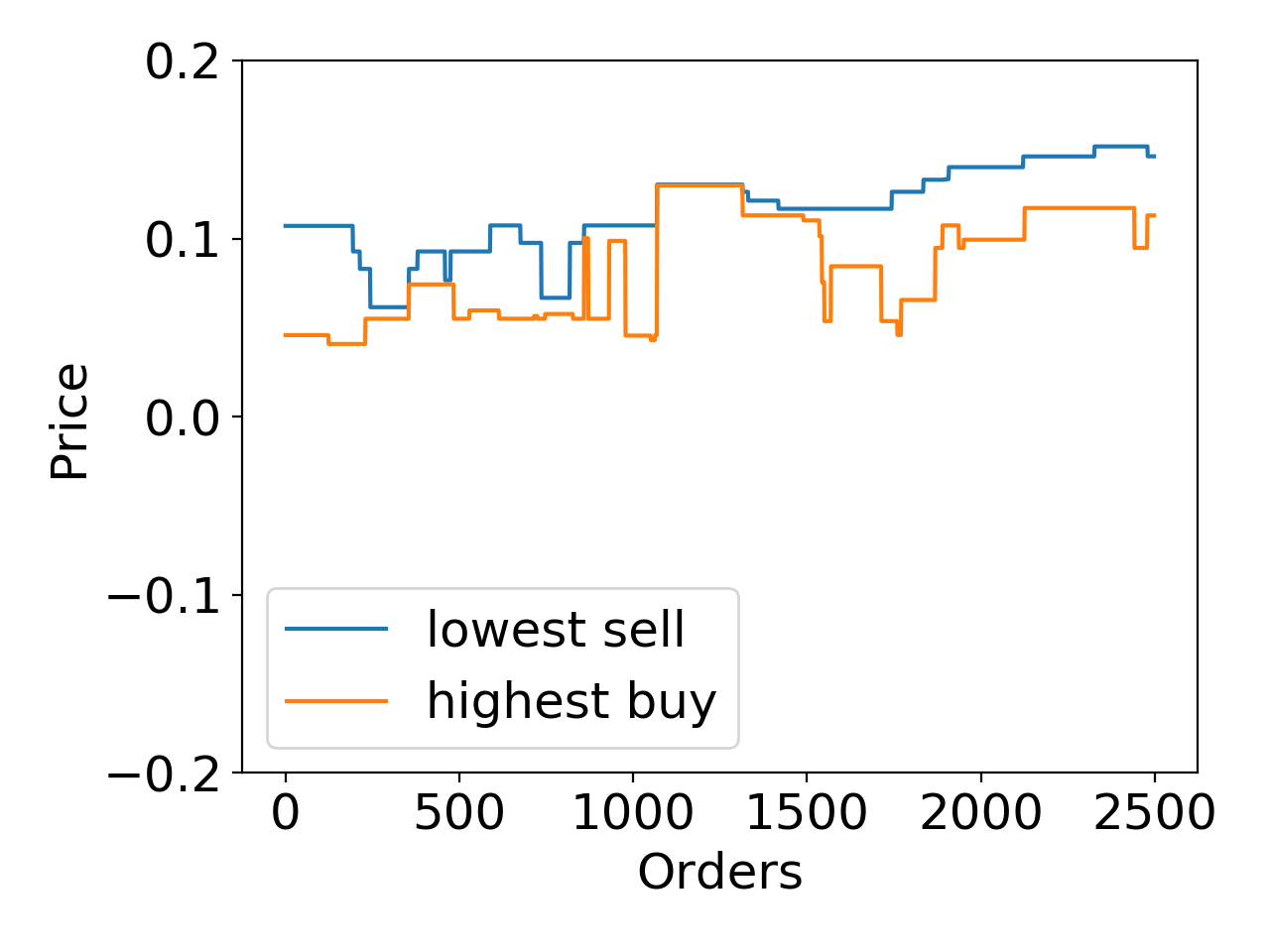}\caption{no CDA network best bid/ask}\label{fig:syn_bestbanocda}\hfill\end{subfigure}
\begin{subfigure}{.24\linewidth}\includegraphics[width=\linewidth]{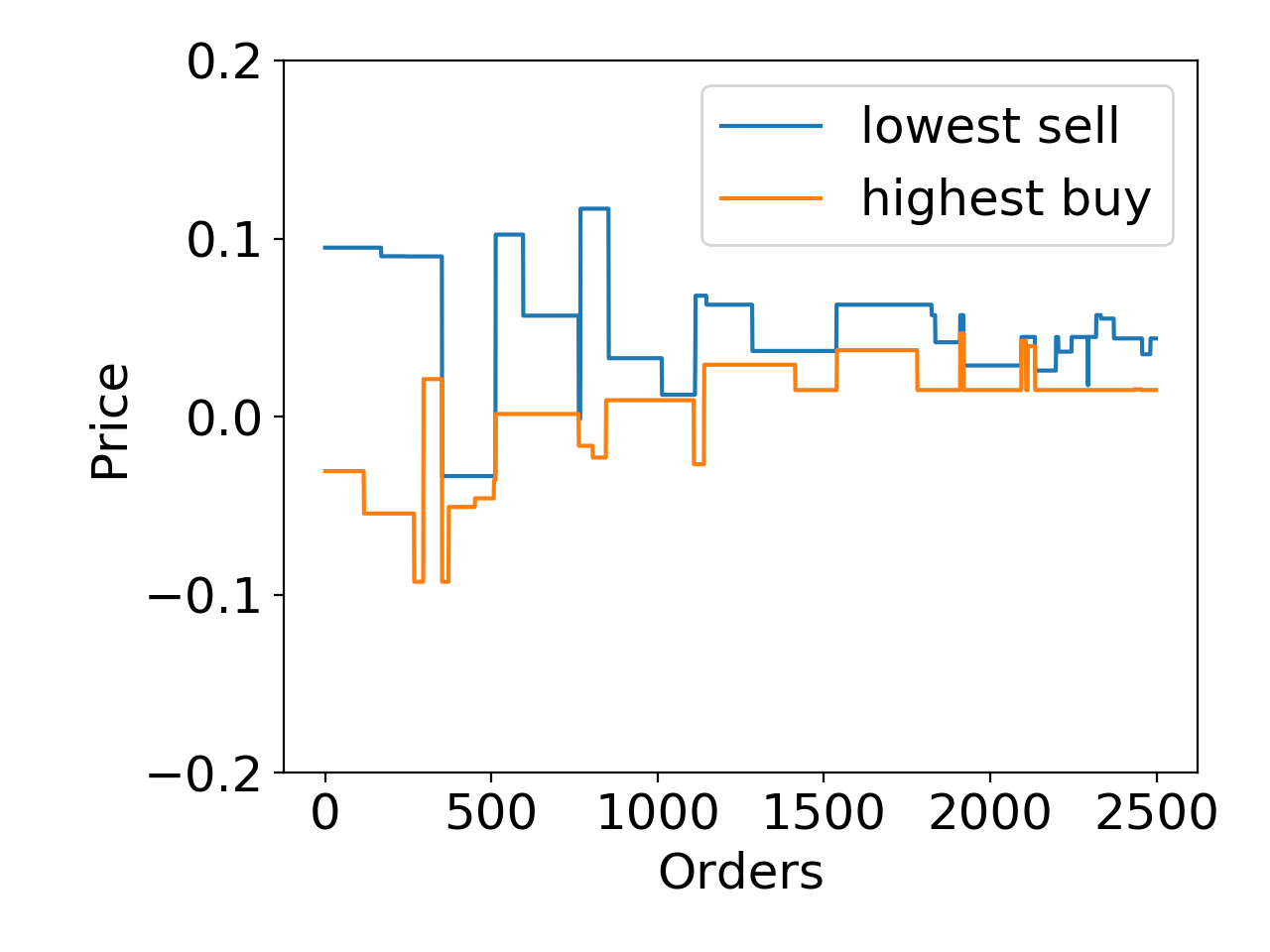}\caption{no order book best bid/ask}\label{fig:syn_bestbanoob}\hfill\end{subfigure}
\hspace*{\fill}\\
\begin{subfigure}{.24\linewidth}\includegraphics[width=\linewidth]{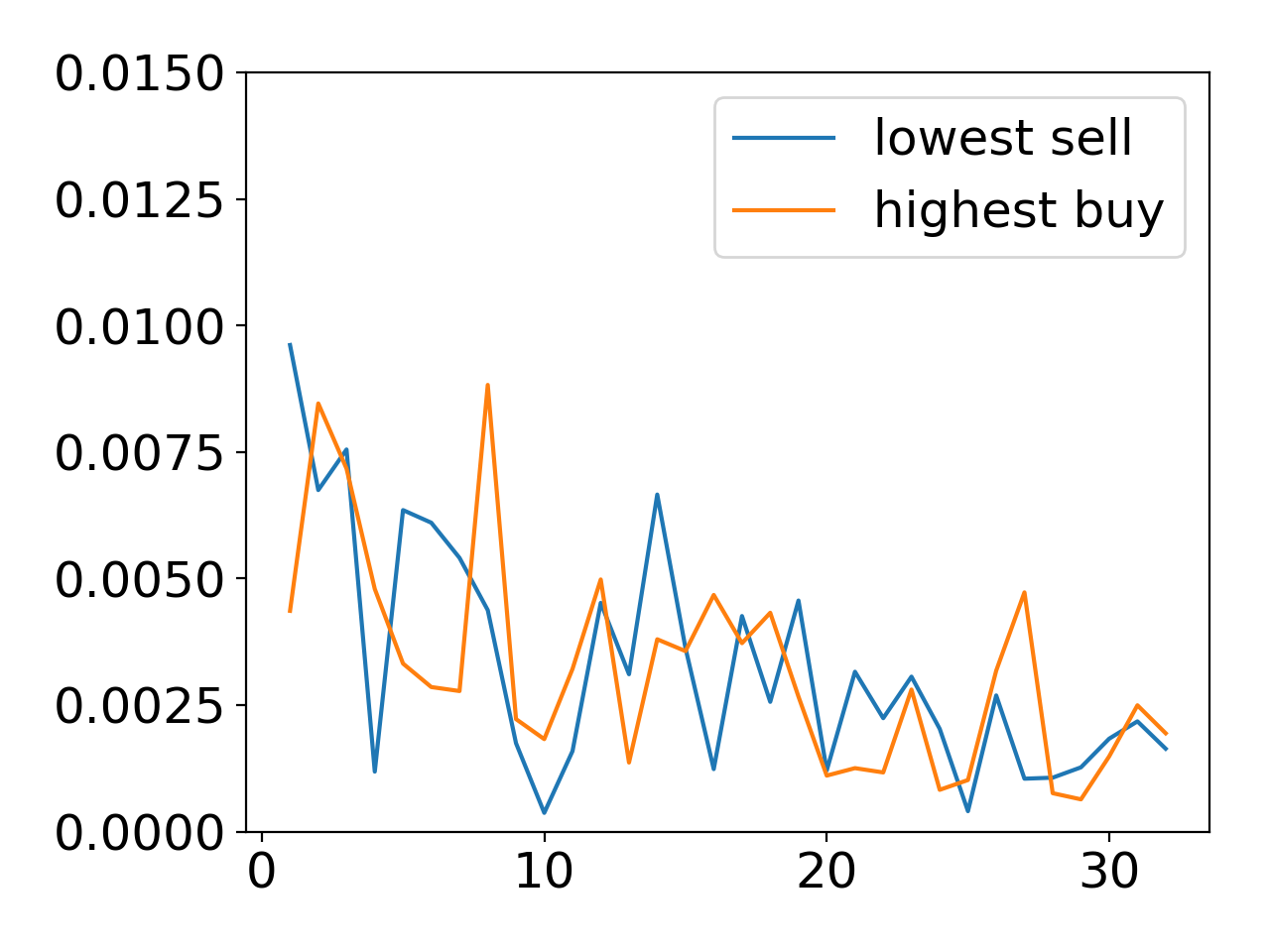}\caption{Synthetic spectral density}\label{fig:syn_spectral}\hfill\end{subfigure}
\begin{subfigure}{.24\linewidth}\includegraphics[width=\linewidth]{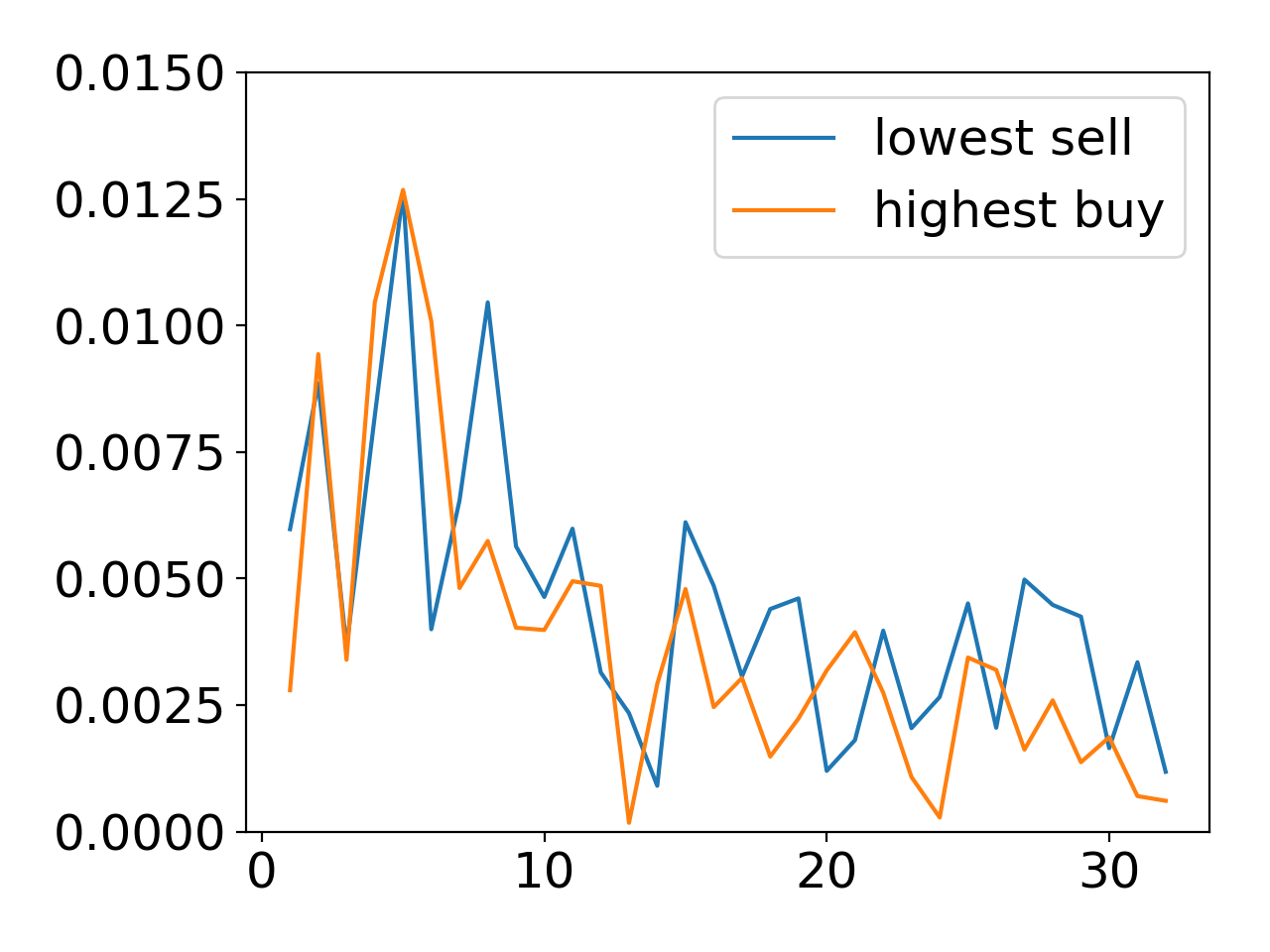}\caption{Stock-GAN spectral density}\label{fig:syn_spectralgen}\hfill\end{subfigure}
\begin{subfigure}{.24\linewidth}\includegraphics[width=\linewidth]{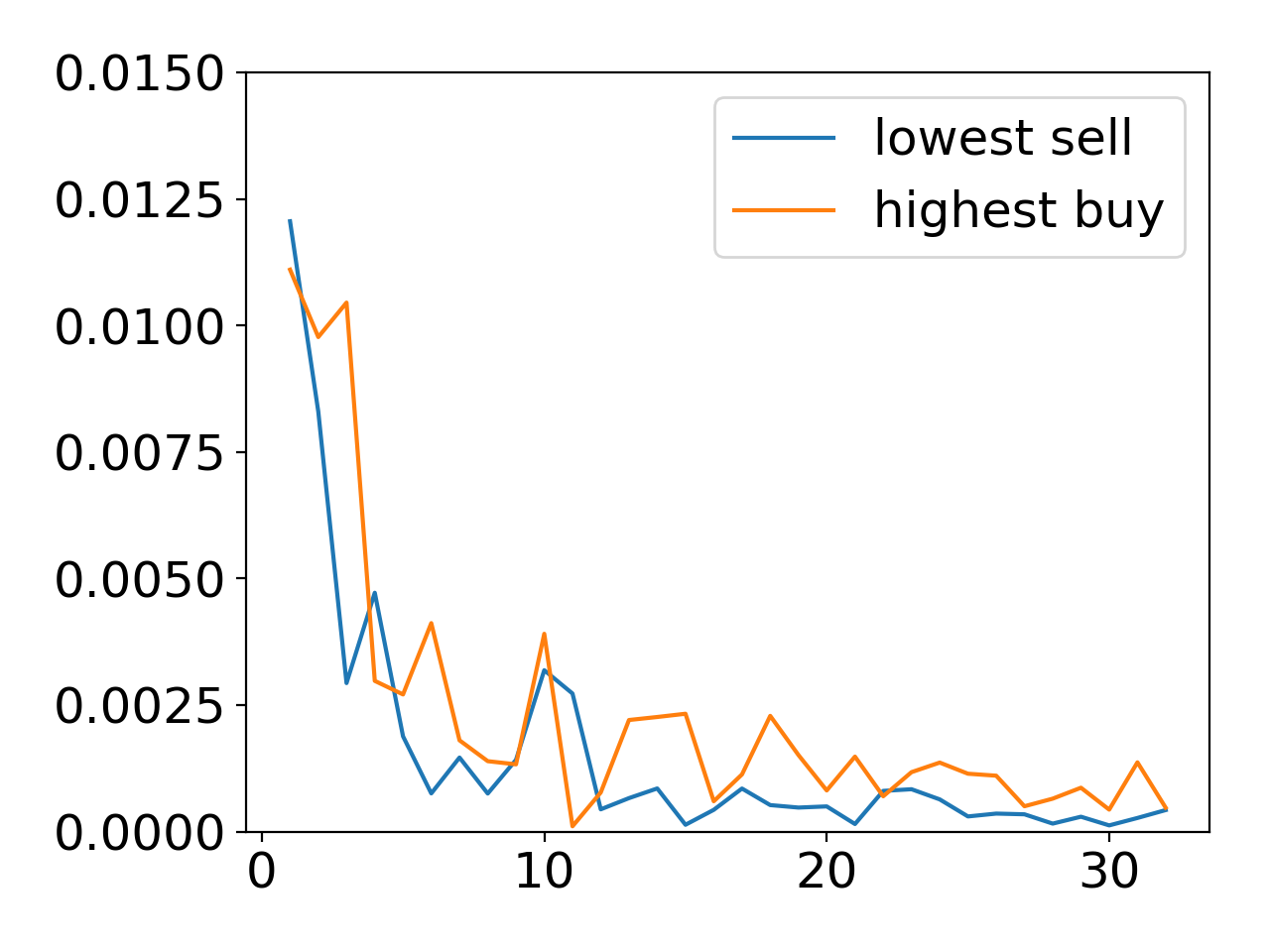}\caption{no CDA spectral density}\label{fig:syn_spectralnocda}\hfill\end{subfigure}
\begin{subfigure}{.24\linewidth}\includegraphics[width=\linewidth]{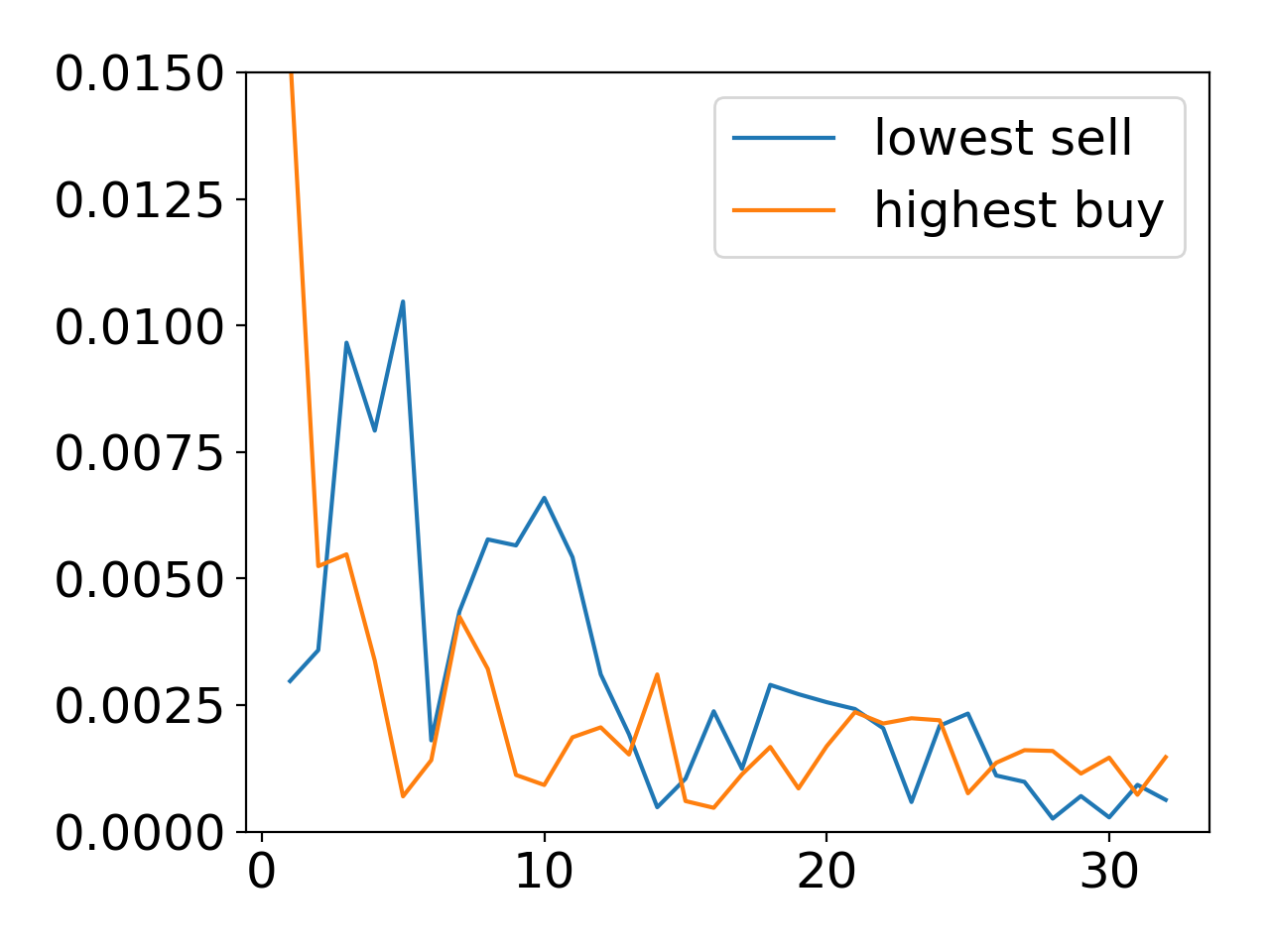}\caption{no order book spectral density}\label{fig:syn_spectralnoob}\hfill\end{subfigure}
\caption{A comparison of different statistics for generated and real synthetic limit orders. Additional results are in appendix.}
\end{figure*}
The best bid and best ask are limit orders themselves and are specified by price and quantity. 
We divide the time in a day into 24 equal intervals and $\Delta \rt$ refers to the index of the interval. A visual representation of $\rvx_i$ is shown in Figure~\ref{1a}.

\subsection{Architecture}
The architecture is shown in Figure~\ref{GAN}. 
We use a conditional WGAN \cite{mirza2014conditional} with both the generator and critic conditioned on a $k$ length history of $\rvx_i$'s and the time interval $\Delta \vt$. We choose $k = 20$. The history is condensed to one vector using a single LSTM layer. This vector and uniform noise of dimension 100 is fed to a  fully connected layer followed by 4 convolution layers. The generator outputs the next $\rvx_i$ and the critic outputs a real number.  Note that when training both generator and critic are fed history from real data, but when the generator executes after training it is fed its own generated data as history. The generator also outputs the best bid and ask as part of $\rvx_i$, which is the output coming out of the CDA network. Recall that the best bid and ask can be inferred deterministically from the current order and the previous best bid and ask (for most orders); we use the CDA network (with frozen weights during GAN training) to output the best bid and best ask; the CDA network serves as a differentiable approximation of the true CDA function. The CDA network has a fully connected layer layer followed by 3 convolutional layers. Its input is a limit order and the current best bid and best ask and the output is the next best bid and best ask. The CDA network is trained separately using the orders and order-book data using a standard mean squared error loss. Appendix~B has the code for generator, critic, and the CDA network that precisely describes the structure, loss, and hyperparameters.

We use the standard WGAN loss with a gradient penalty term \cite{gulrajani2017improved}. The critic is trained $100$ times in each iteration. The \emph{notable} part in constructing the training data is that for each of 64 data points in a mini-batch the sequence of orders chosen (including history) is far away from any other sequence in that mini-batch. This is to break the dependence among data points for the history dependent stock market data. We make this mathematically precise next.

\subsection{Mathematical Characterization of Stock-GAN}
We show how general stochastic process view of limit order generation provides an interpretation of the distribution that the generator that Stock-GAN is learning. 
Recall that a stochastic process is a collection of random variables indexed by a set of numbers. We view the stock market orders for a given chunk of time of day $\Delta \rt$ as a collection of vector valued random variable $\{\rvx_i\}_{i\in N}$ indexed by the limit order sequence number in $N = \{1, \ldots, n\}$, where $n$ is the maximum number of limit orders that can possibly show up in any $\Delta t$ time interval. 
Following the terminology for stochastic processes, the above process is discrete time and discrete space (discrete time here refers to the discreteness of the index set $N$)\@.

The $k$ length history we use implies a finite history dependence of the current output $\rvx_i$, that is, $P(\rvx_i\mid \rvx_{i-1},\ldots, \Delta \rt) = P(\rvx_i\mid \rvx_{i-1},\ldots, \rvx_{i-m}, \Delta \rt)$ for some $m$. Such dependence is justified by the observation that recent orders mostly determine the transactions and transaction price in the market as orders that have been in the market for long either get transacted or canceled. 
Further, the best bid and best ask serves as an (approximate) sufficient statistic for events beyond the history length $m$. While this process is not a Markov chain (MC), it forms what is known as a higher order MC, which implies that the process given by $\rvy_i = (\rvx_i, \ldots, \rvx_{i-m+1})$ is a MC for any given time interval $\Delta \rt$. We assume that this chain formed by $\rvy_i$ has a stationary distribution (i.e., it is irreducible and positive recurrent). A MC is a \emph{stationary stochastic process} if it starts with its stationary distribution. After some initial mixing time, the MC does reach its stationary distribution, thus, we assume that the process is stationary by throwing away some initial data for the day. Also, for the jumps across two time intervals $\Delta \rt$, we assume the change in stationary distribution is small and hence the mixing happens very quickly. A stationary process means that $P(\rvx_i, \ldots, \rvx_{i-m+1}\mid \Delta \rt)$ has the same distribution for any $i$.
In practice we do not know $m$. However, we assume that our choice $k$ satisfies $k+1 > m$, and then it is straightforward to check that $\rvy_t = (\rvx_i, \ldots, \rvx_{i-k})$ is a MC and the claims above hold with $m-1$ replaced by $k$. Note that unlike simple stochastic processes for transaction prices (or fundamental value of a stock) used in finance literature, such as the mean reverting Ornstein-Uhlenbeck process, our stochastic process of market order has a complex random variable per time step and cannot be described in a closed form. Hence, we use a neural network to learn this complex stochastic process.

Given the above stochastic process view, we show that the generator aims to learn the real conditional distribution $P_r(\rvx_i\mid \rvx_{i-1},\ldots, \rvx_{i-k}, \Delta \rt)$. We use the subscript $r$ to refer to real distributions and the subscript $g$ to refer to generated distributions. The real data $\vx_1, \vx_2, \ldots$ is a realization of the stochastic process. It is worth noting that even though $P(\rvx_i, \ldots, \rvx_{i-k}\mid \Delta \rt)$ has the same distribution for any $i$, the realized real data sequence $\vx_i, \ldots \vx_{i-k}$ is correlated with any overlapping sequnce $\vx_{i+k'}, \ldots \vx_{i-k+k'}$ for $k \geq k' \geq -k$. Our training data points are sequences $\vx_i, \ldots \vx_{i-k}$ and as stated earlier we make sure that the sequences in a batch are sufficiently far apart. In light of the interpretation above, this ensures independence of data points within a batch.

\textbf{Critic interpretation}: When fed real data, the critic can be seen as a function $c_w$ of the realized data $\vs_i=(\vx_i, \ldots , \vx_{i-k}, \Delta \vt)$, where $w$ are the weights of the critic network.  
As argued earlier, samples in a batch that are chosen from real data that are spaced at least $k$ apart are i.i.d.  samples of $P_r$. 
Then for $m$ samples fed to the critic, $\frac{1}{m}\sum_{i=1}^{m} c_w(\vs_i)$ estimates $E_{\rvs \sim P_r}(c_w(\rvs))$. When fed generated data (with the ten price levels determined from the output order and previous ten levels), by similar reasoning $\frac{1}{m}\sum_{i=1}^{m} c_w(\vs_i)$ estimates $E_{\rvs \sim P_g}(c_w(\rvs))$ when the samples are sufficiently apart (recall that the history is always real data). Thus, the critic computes the Wasserstein distance between the joint distributions $P_r(\rvx_i, \ldots, \rvx_{i-k}, \Delta \rt)$ and $P_g(\rvx_i, \ldots, \rvx_{i-k}, \Delta \rt)$.

\textbf{Generator interpretation}: The generator learns the conditional distribution $P_g(\rvx_i\mid \rvx_{i-1}, \ldots, \rvx_{i-k}, \Delta \rt)$. Along with the real history that is fed during training, the generator represents the distribution $P_g(\rvx_i, \ldots, \rvx_{i-k}, \Delta \rt) = P_g(\rvx_i\mid \rvx_{i-1}, \ldots, \rvx_{i-k}, \Delta\rt)P_r(\rvx_{i-1}, \ldots, \rvx_{i-k}, \Delta\rt)$. 

\section{Experimental Results}
Evaluating generative models is an inherently challenging task, even in the well-established domain of image generation \cite{BORJI2018}. 
To the best of our knowledge, we are the first to generate limit order streams in stock market that is calibrated to real data and as part of our contribution we propose to measure the quality of generated data using five statistics. These statistics capture various aspects of order streams observed in stock markets that are often studied in finance literature.

Our five proposed statistics are 
\begin{enumerate}
\item Price: Distribution over price for the day's limit orders, by order type.
\item Quantity: Distribution over quantity for the day's limit orders, by order type.
\item Inter-arrival time: Distribution over inter-arrival duration for the day's limit orders, by order type.
\item Intensity evolution: Number of orders for consecutive $T$-second chunks of time.
\item Best bid/ask evolution: Changes in the best bid and ask over time as new orders arrive.
\end{enumerate}
For each of these statistics, we also present various quantitative numbers to measure the quality. Due to lack of space, in the main paper the results for price, quantity, inter-arrival distributions are shown only for buy orders. The results for the other types are similar to buy type results and presented in the appendix.

\begin{figure}[t!]
\centering
\begin{subfigure}{.75\linewidth}{\includegraphics[width=\linewidth]{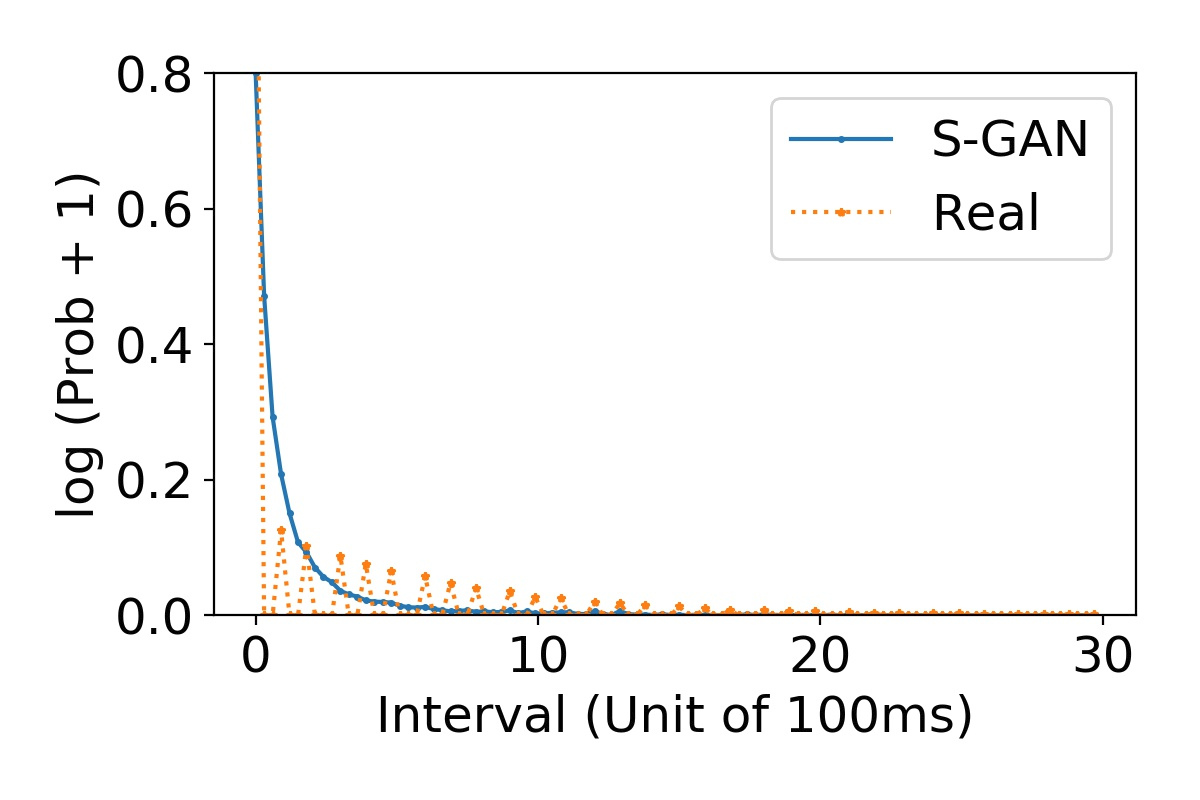}\phantomsubcaption\label{fig:syn_interarrival_1}}\hfill\end{subfigure}
\begin{subfigure}{.75\linewidth}{\includegraphics[width=\linewidth]{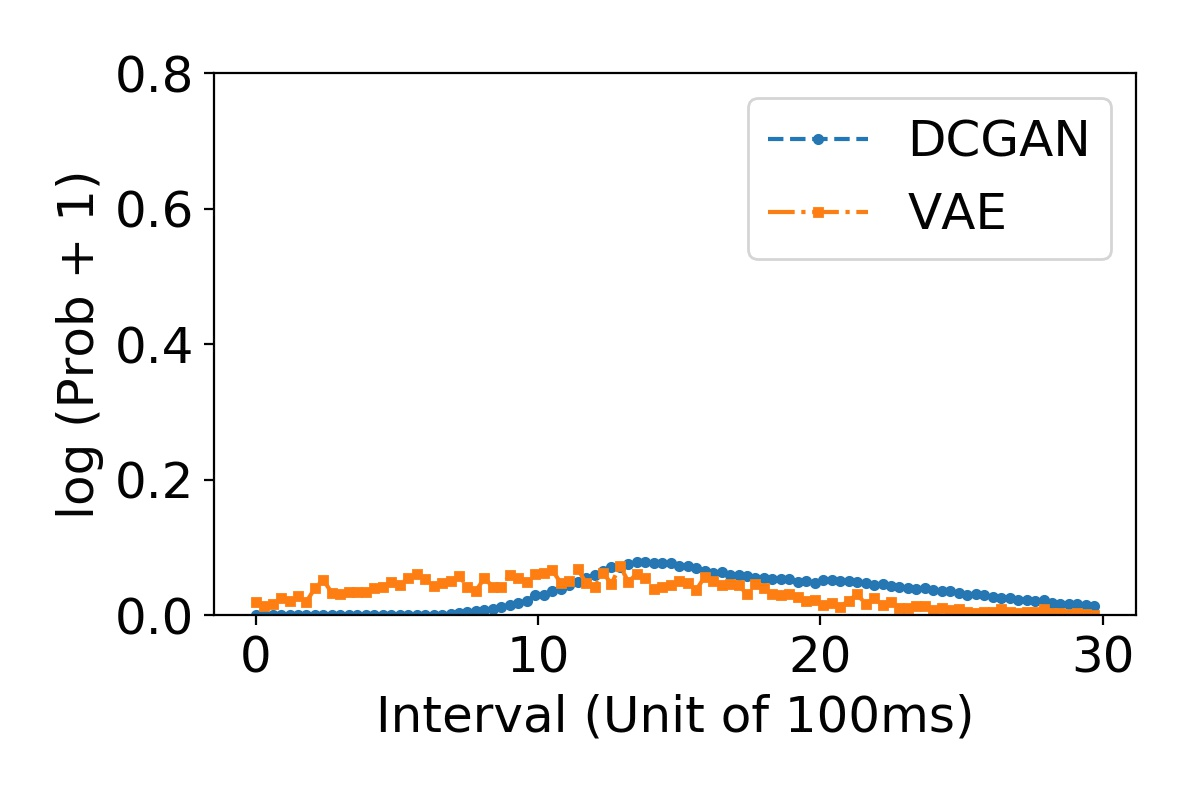}\phantomsubcaption\label{fig:syn_interarrival_2}}\hfill\end{subfigure}
\caption{Synthetic inter-arrival distribution}
\end{figure}

\subsection{Synthetic Data}
\begin{figure*}[t!]
\begin{subfigure}{.33\linewidth}\includegraphics[width=\linewidth]{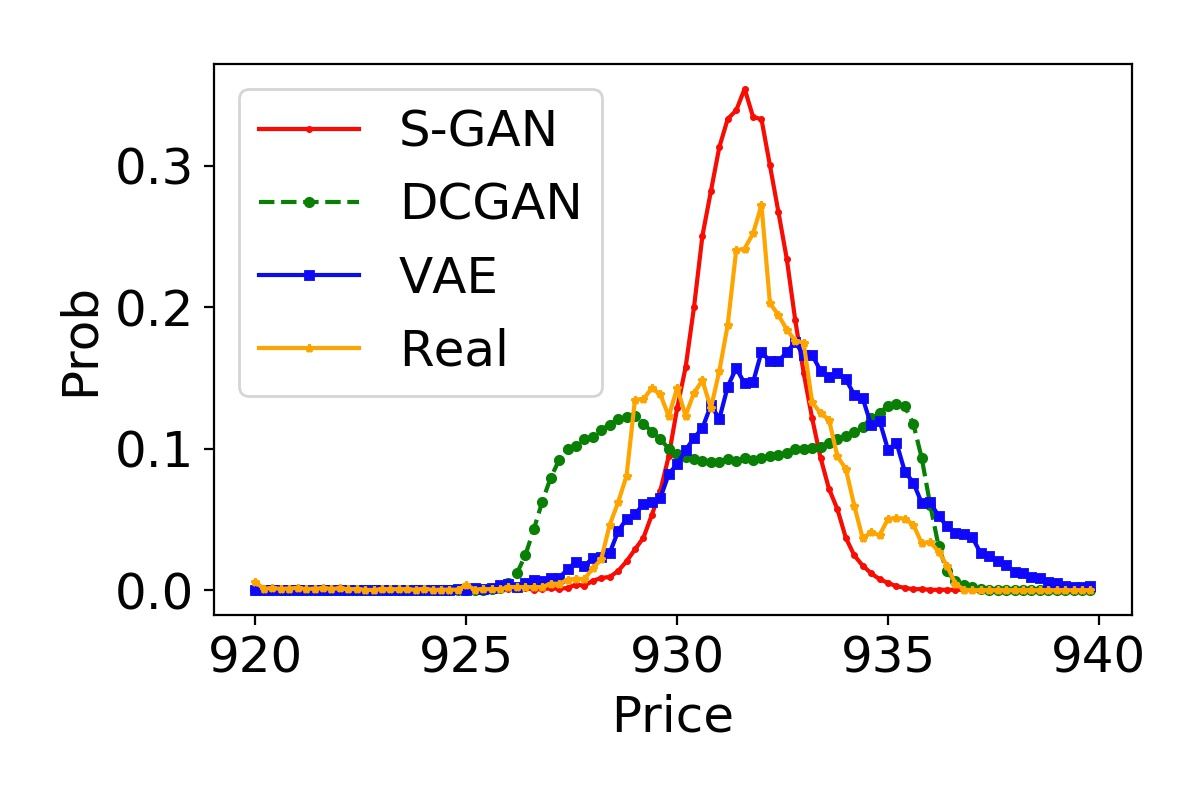}\caption{GOOG price distribution}\label{fig:goog_price}\hfill\end{subfigure}
\begin{subfigure}{.33\linewidth}\includegraphics[width=\linewidth]{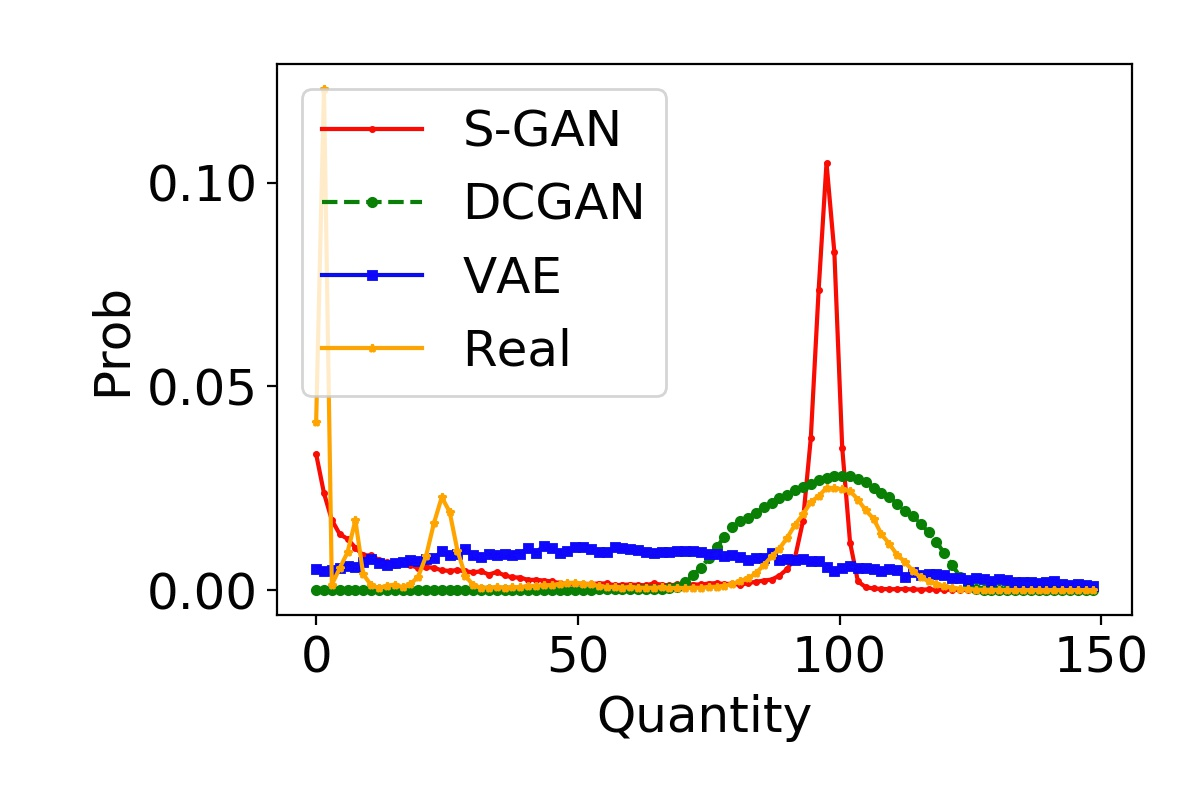}\caption{GOOG quantity distribution}\label{fig:goog_qnt}\hfill\end{subfigure}
\begin{subfigure}{.33\linewidth}\includegraphics[width=\linewidth]{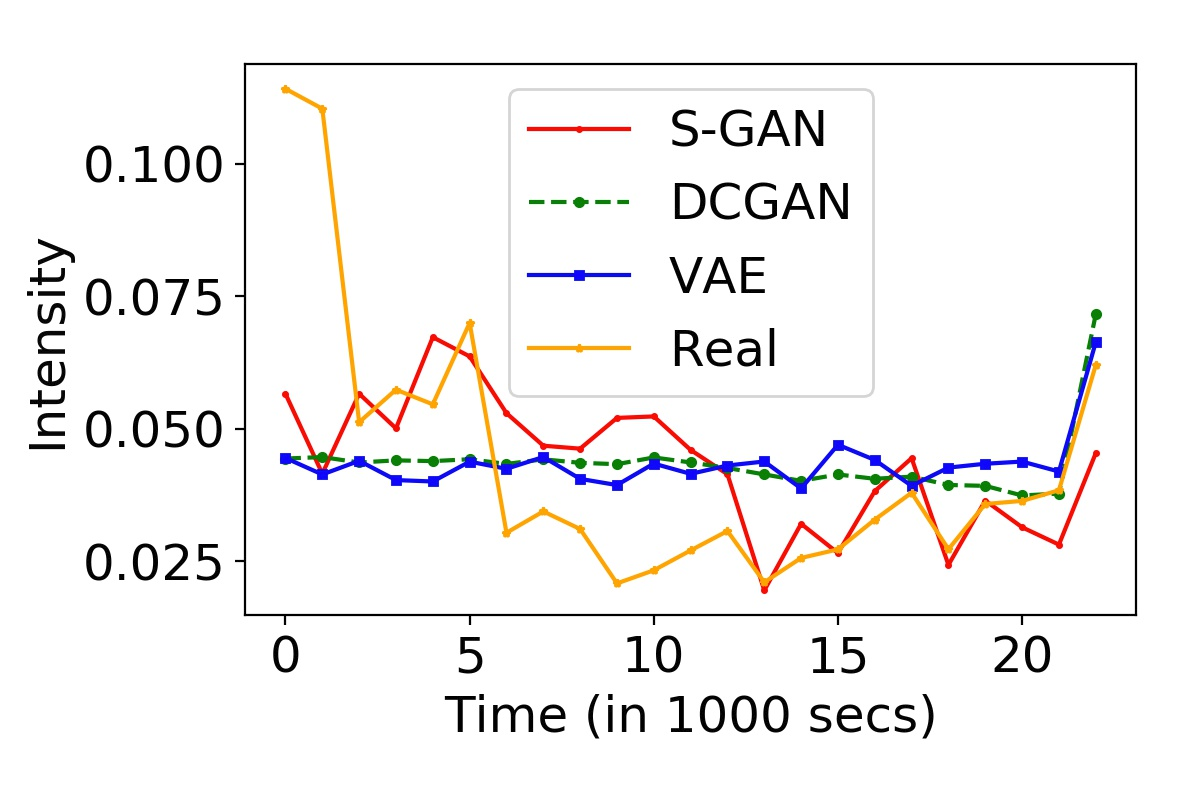}\caption{GOOG intensity}\label{fig:goog_int}\hfill\end{subfigure}
\hspace*{\fill}
\begin{subfigure}{.24\linewidth}\includegraphics[width=\linewidth]{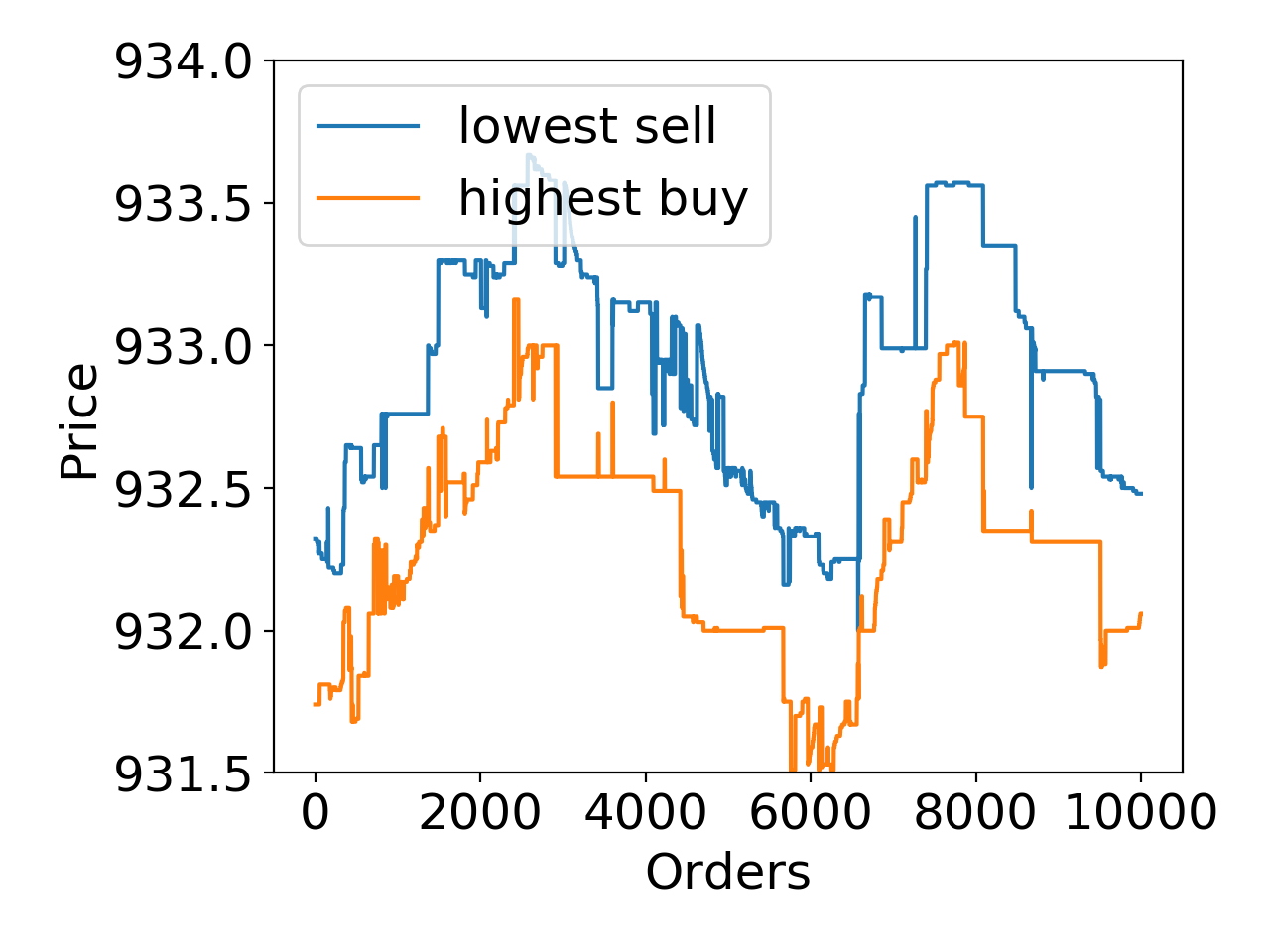}\caption{GOOG best bid/ask}\label{fig:goog_bestba}\hfill\end{subfigure}
\begin{subfigure}{.24\linewidth}\includegraphics[width=\linewidth]{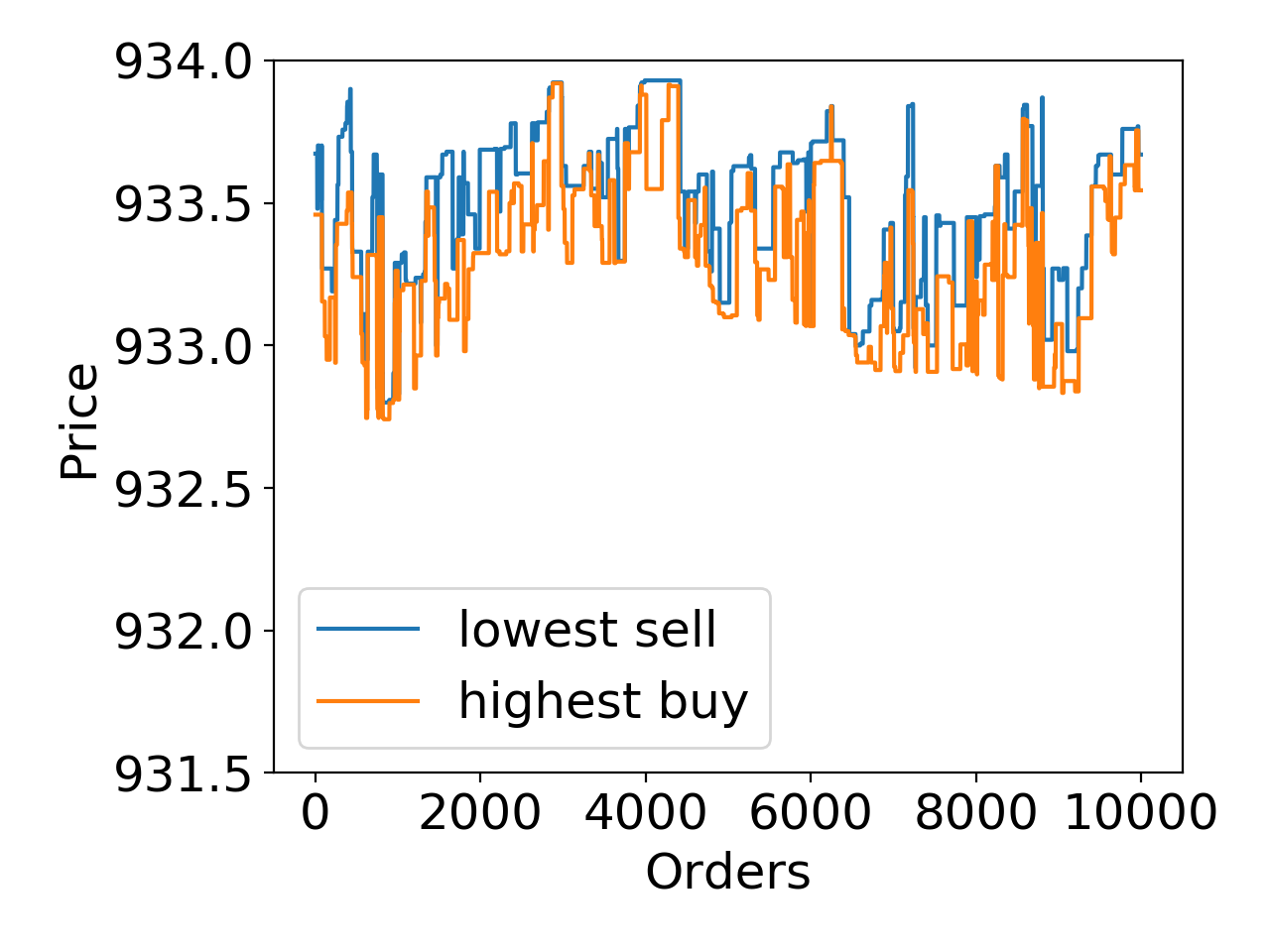}\caption{Stock-GAN best bid/ask}\label{fig:goog_bestbagen}\hfill\end{subfigure}
\begin{subfigure}{.24\linewidth}\includegraphics[width=\linewidth]{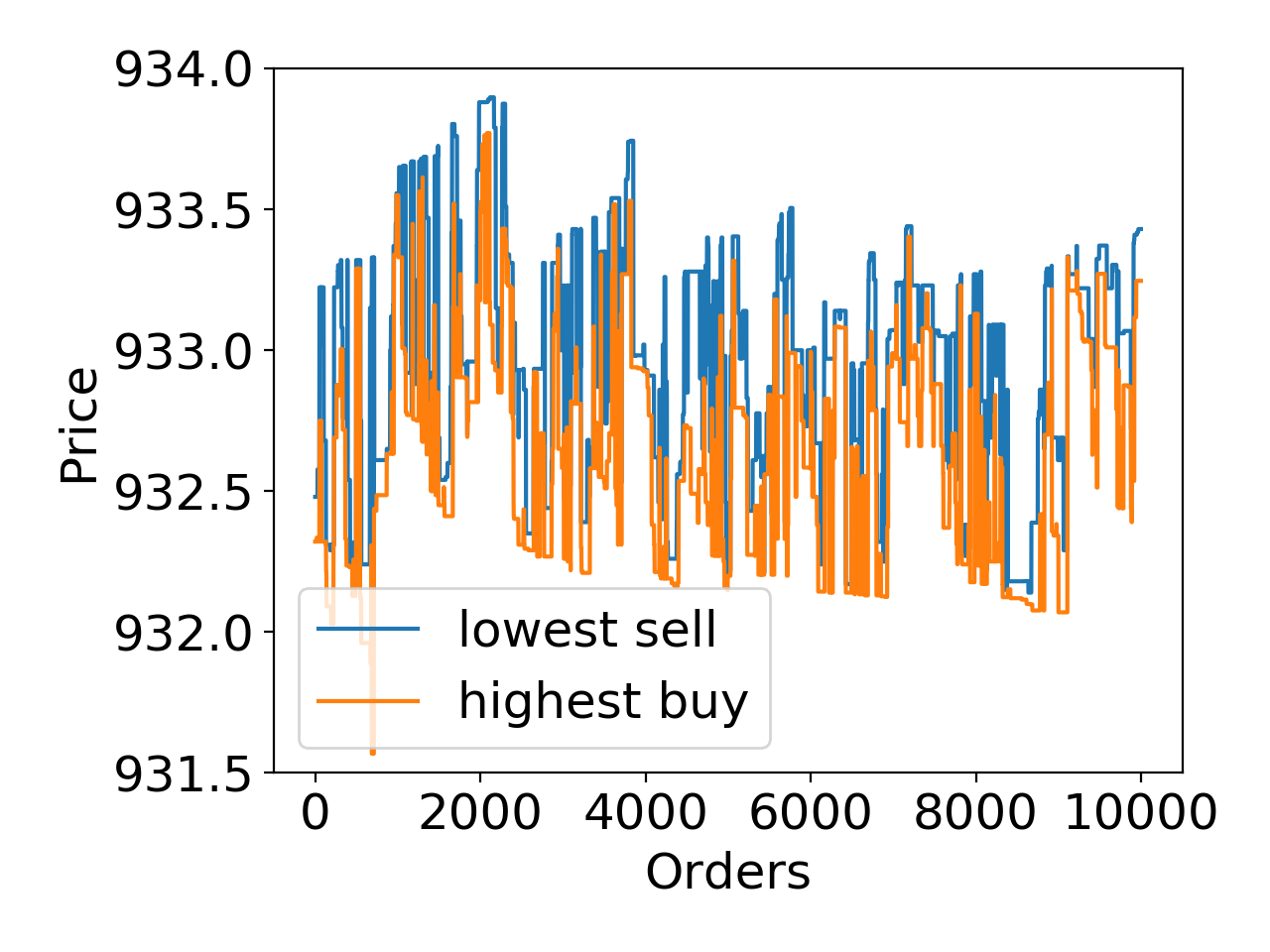}\caption{no CDA network best bid/ask}\label{fig:goog_bestbanocda}\hfill\end{subfigure}
\begin{subfigure}{.24\linewidth}\includegraphics[width=\linewidth]{ablation_goog_predict_goog_no_cda.png}\caption{no order book best bid/ask}\label{fig:goog_bestbanoob}\hfill\end{subfigure}
\hspace*{\fill}
\begin{subfigure}{.24\linewidth}\includegraphics[width=\linewidth]{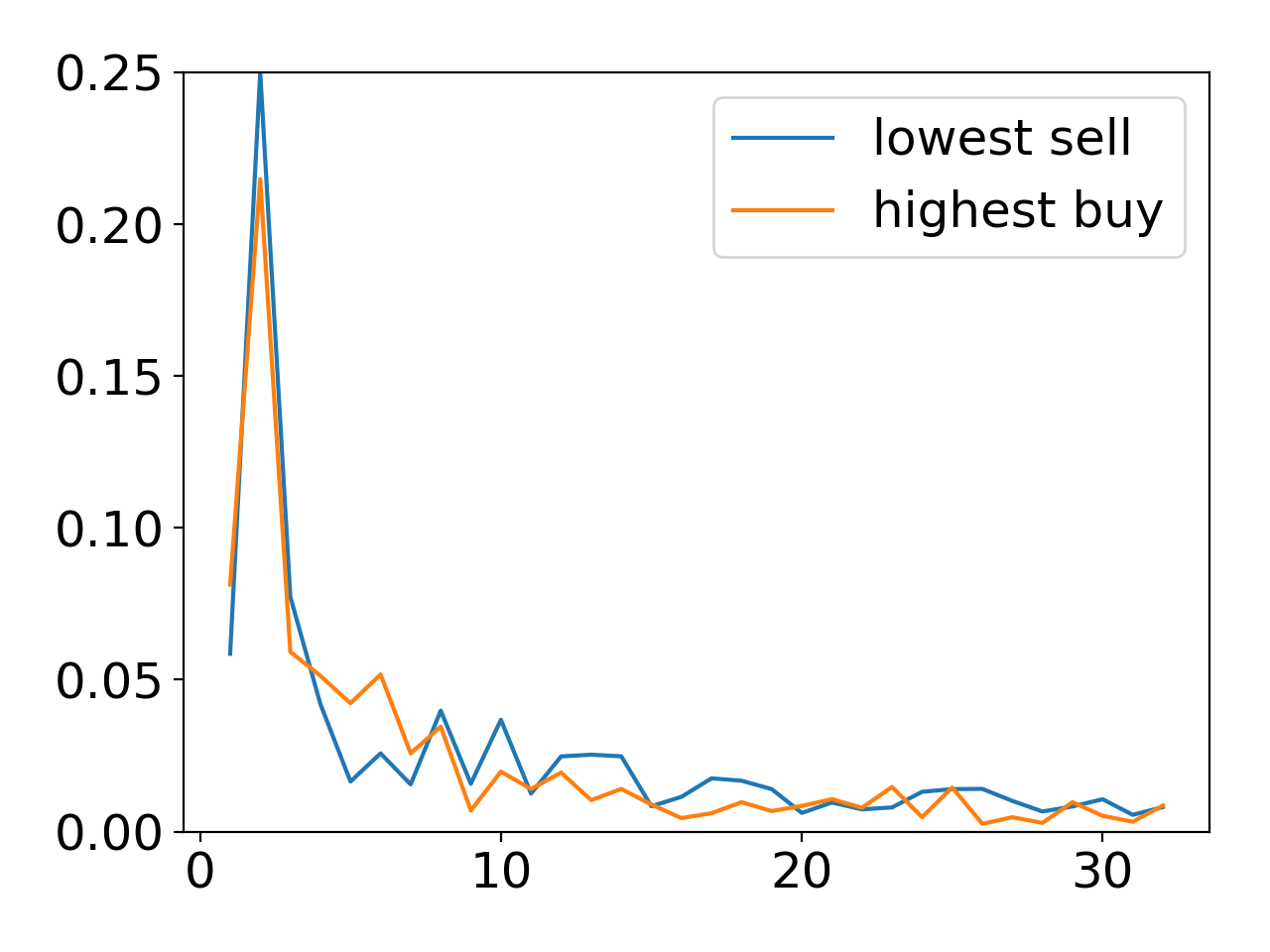}\caption{GOOG spectral bid/ask}\label{fig:goog_spectral}\hfill\end{subfigure}
\begin{subfigure}{.24\linewidth}\includegraphics[width=\linewidth]{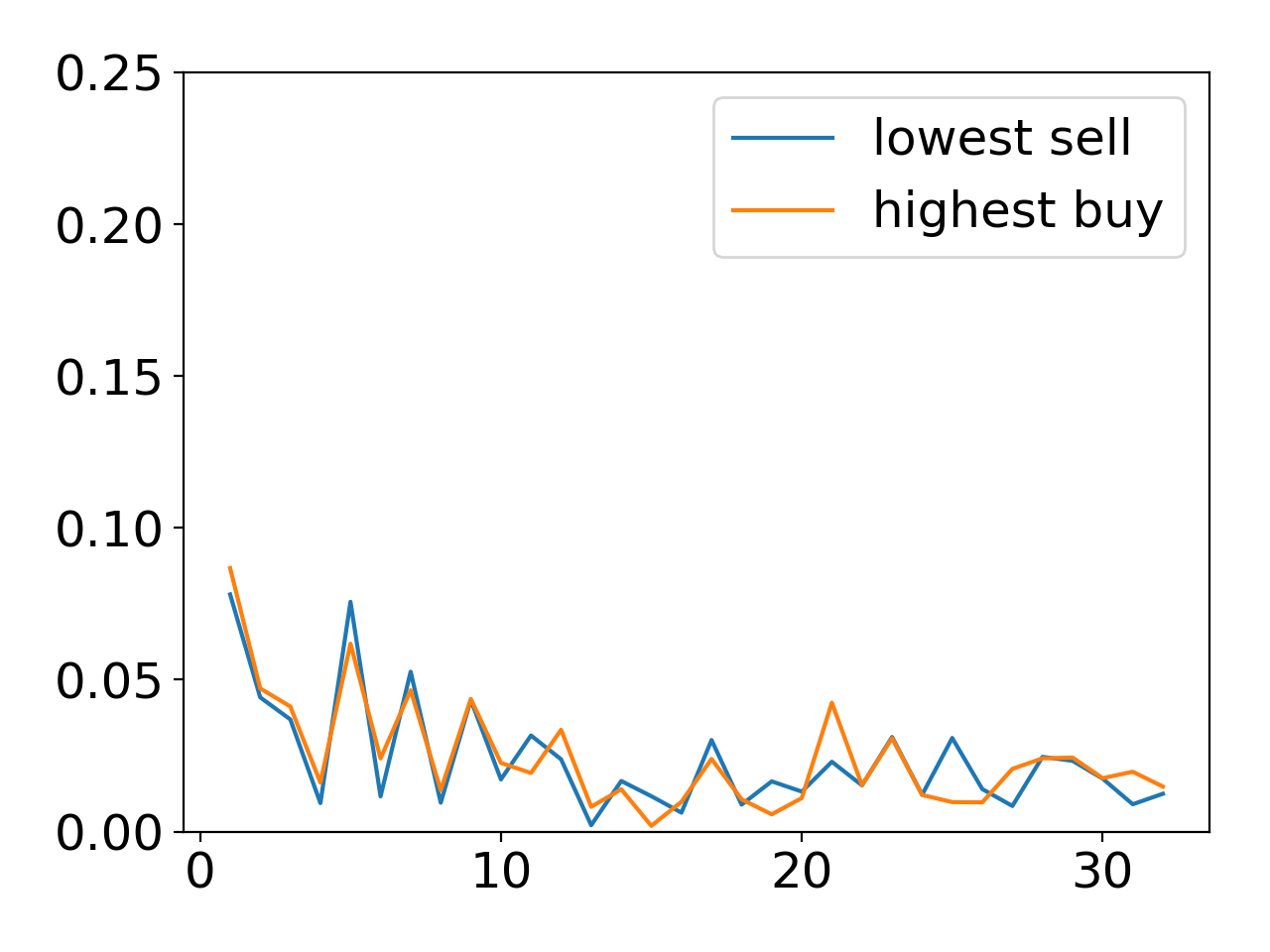}\caption{Stock-GAN spectral density}\label{fig:goog_spectralgen}\hfill\end{subfigure}
\begin{subfigure}{.24\linewidth}\includegraphics[width=\linewidth]{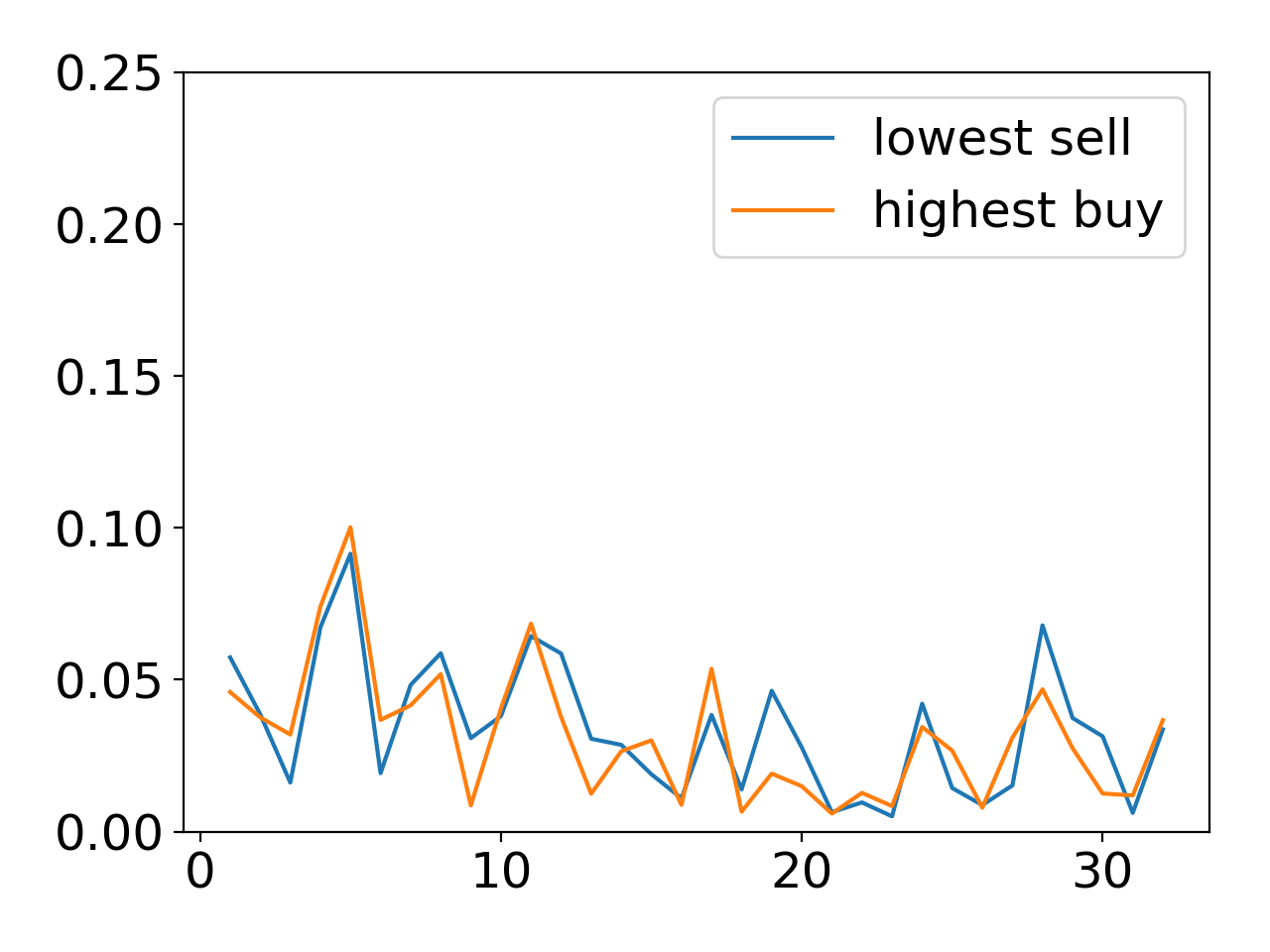}\caption{no CDA spectral density}\label{fig:goog_spectralnocda}\hfill\end{subfigure}
\begin{subfigure}{.24\linewidth}\includegraphics[width=\linewidth]{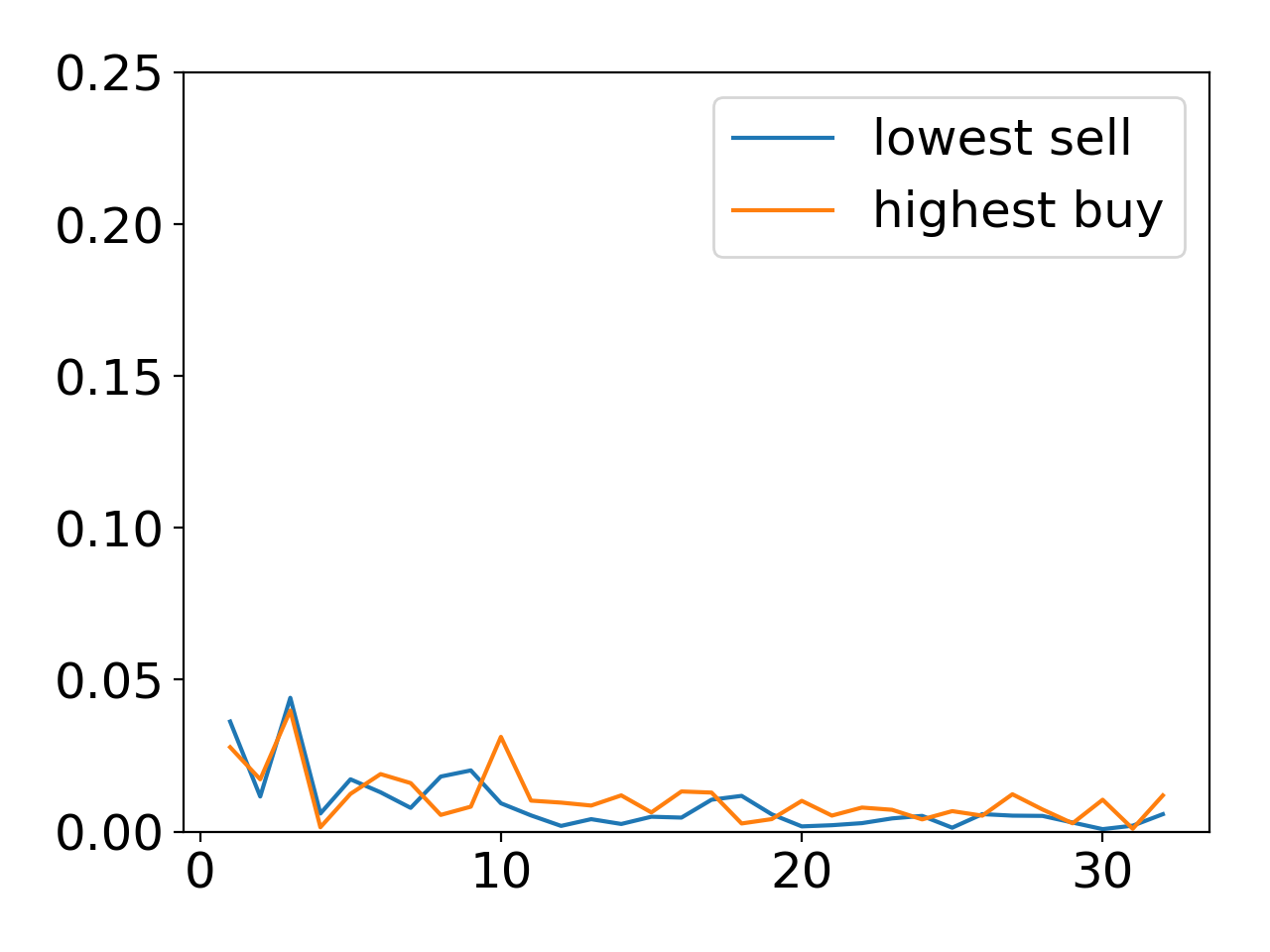}\caption{no order book spectral density}\label{fig:goog_spectralnoob}\hfill\end{subfigure}
\caption{A comparison of different statistics for generated and real GOOG limit orders. Additional results are in appendix.}
\label{fig:goog}
\end{figure*}

We first evaluate Stock-GAN on synthetic orders generated from an agent-based market simulator. 
Previously adopted to study a variety of issues in financial markets (e.g., market making \cite{Wah17ww} and manipulation \cite{Wang18vw}), the simulator captures stylized facts of the complex financial market with specified stochastic processes and distributions \cite{wellman2017strategic}. 
However, the simulator is still very basic and quite far from real market data. For example, fundamental valuation shocks are generated from a fixed Gaussian distribution (Figure~\ref{fig:syn_price}) and quantity is always 1 (Figure~\ref{fig:syn_qnt}), whereas the real market data distributions can be seen to be quite non-smooth (Figures~\ref{fig:goog_price}-~\ref{fig:goog_int}). Thus, we use the output of this basic simulator as our synthetic data (which we call as real in results below).
We use about 300,000 orders generated by the simulator as our synthetic data. These orders are generated over a horizon of 1000 seconds, but the actual horizon length is not important for synthetic data as it can be scaled arbitrarily. The price output by the simulator is normalized to $[-1,1]$, which is the reason for negative prices in the synthetic data.

\textbf{Stock-GAN and baselines}: Our first results show the performance of Stock-GAN (S-GAN in graphs) and compares it to baselines, namely to a recurrent variational autoencoder~\cite{chung2015recurrent} (VAE) and the same network as ours, except using a DCGAN~\cite{radford2015unsupervised} instead of WGAN. We show results for price distribution (Figure~\ref{fig:syn_price}), quantity distribution (Figure~\ref{fig:syn_qnt}),  and inter-arrival distribution (Figure~\ref{fig:syn_interarrival_1}, \ref{fig:syn_interarrival_2}---shown in two larger graphs for clarity). The results show that VAE and DCGAN produce distributions far from the real one. We capture these differences quantitatively using the Kolmogorov-Smirnoff (KS) distance (Table~\ref{KSsyn}).
\begin{table}[t]
\centering
\resizebox{.9\columnwidth}{!}{
\begin{tabular}{l c c c}  
\toprule
   & Real,S-GAN & Real,VAE & Real,DCGAN  \\
\midrule
Price    & 0.108 & 0.502 & 0.284         \\
Inter-arrival  & 0.18 & 0.756 & 0.923    \\
\bottomrule
\end{tabular}
}
\caption{KS distances against real (synthetic)} \label{KSsyn}
\end{table}

\begin{figure}[t!]
\centering
\begin{subfigure}{.75\linewidth}\includegraphics[width=\linewidth]{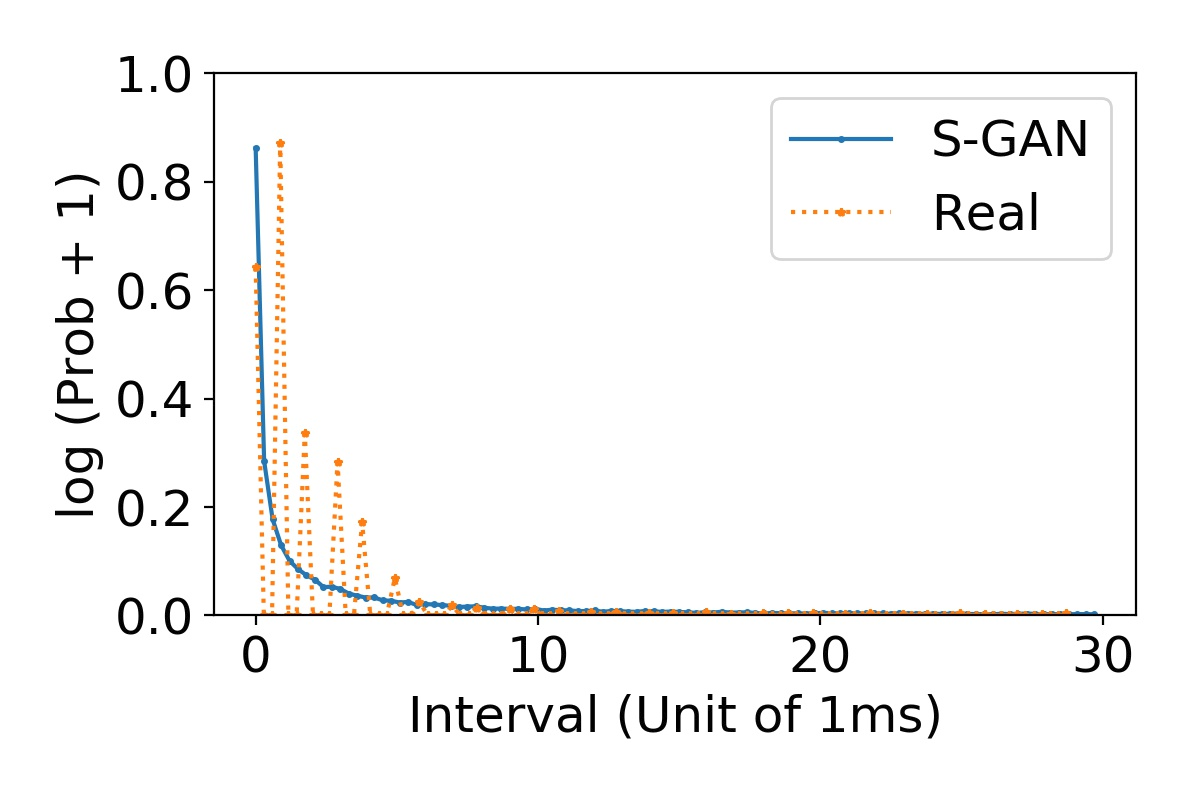}\phantomsubcaption\label{fig:goog_interarrival_1}\hfill\end{subfigure}
\begin{subfigure}{.75\linewidth}\includegraphics[width=\linewidth]{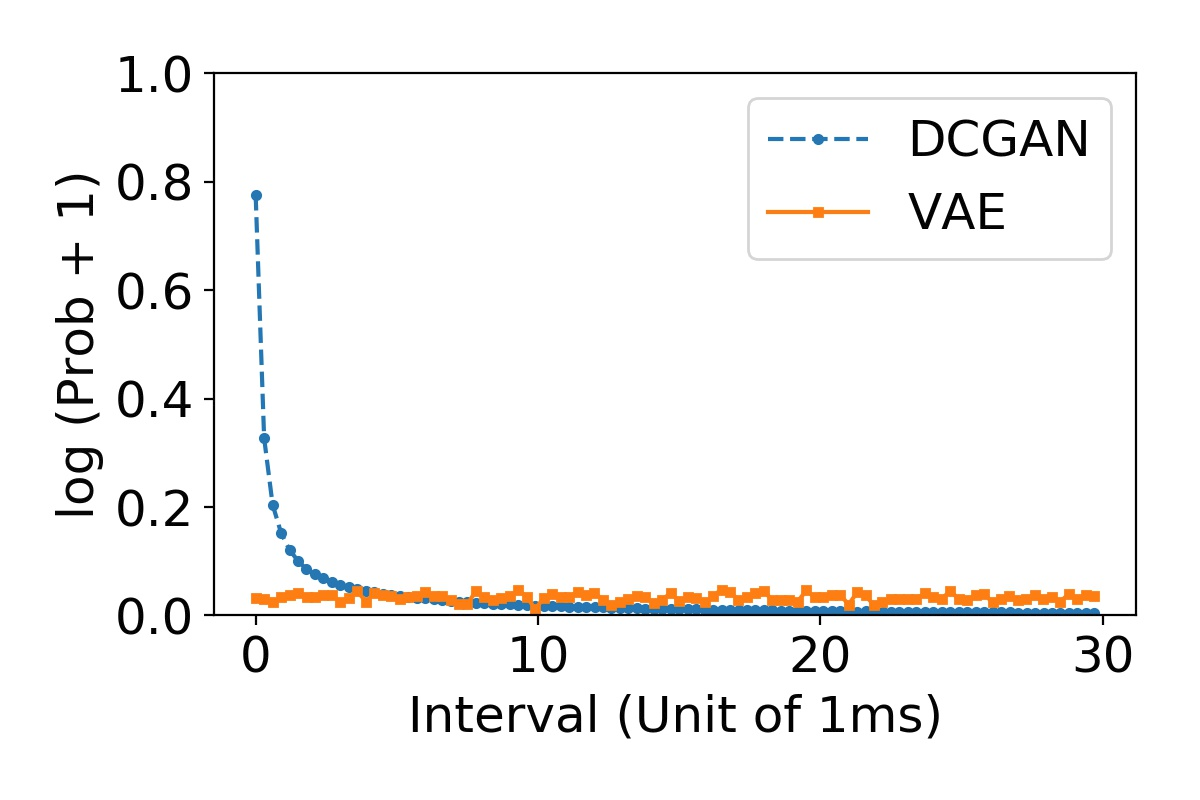}\phantomsubcaption\label{fig:goog_interarrival_2}\hfill\end{subfigure}
\caption{GOOG inter-arrival distribution}
\end{figure}

The KS distance is always in $[0,1]$. We skip the KS distance between quantity, which is always trivially one in the synthetic data.
The much smaller KS distance between real and Stock-GAN supports our claim of better performance of Stock-GAN compared to VAE and DCGAN.

For intensity, we choose $T=100$ seconds sized chunks of time and measure intensity as the number of orders in each chunk divided by the total number of orders.
Figure~\ref{fig:syn_int} shows that VAE  completely fails to match the real (synthetic) data intensity. DCGAN has the same flat intensity throughout and again failing to match the real data intensity completely. In contrast, Stock-GAN matches the real data intensity very closely.

\textbf{Ablation}: The real and Stock-GAN generated best bid/ask evolutions are in Figure~\ref{fig:syn_bestba} and~\ref{fig:syn_bestbagen} respectively. We perform two ablation experiments, one by removing the CDA network (no cda) and one by removing order-book information (no ob), shown in Figures\ \ref{fig:syn_bestbanocda} and~\ref{fig:syn_bestbanoob} respectively. Differences can be seen in best bid/ask means for no cda and no ob compared to the real and Stock-GAN results, but the quantitative distinction is in the spectral densities for these time series shown in Figures\ \ref{fig:syn_spectral}--\ref{fig:syn_spectralnoob}. The spectral density of a time series is the magnitude of each frequency component in the Fourier transform of the time series. The spectral density figures shows the frequency component magnitude for every frequency on the x-axis, which is a quantitative means of comparing two time series. It can be seen that no cda and no ob have much fewer higher frequency components as compared to synthetic spectral density, which can also be seen by the smoother time variation in Figures \ref{fig:syn_bestbanocda} and~\ref{fig:syn_bestbanoob}. Stock-GAN's and the synthetic spectral density match more closely.

\subsection{Real Data}
We obtained real limit-order streams from OneMarketData, who provided access to their OneTick database for selected time periods and stocks.
The provided data streams comprise order submissions and cancellations at millisecond granularity.
In experiments, we evaluate the performance of Stock-GAN on a large capitalization stock, Alphabet Inc (GOOG). We also tried a small capitalization stock Patriot National (PN).
After pre-processing, the PN daily order stream has about 20,000 orders and GOOG has about 230,000. Hence, naturally PN is not a good fit for learning using data hungry neural networks and our results for PN (shown in appendix) validate this claim. 

Relative to synthetic data, the real market data is very noisy including many orders at extreme prices far from the range where transactions occur.
Since our interest is primarily on behavior that can affect market outcomes, we focus on orders in the relevant range near the best bid and ask.
Specifically, in a preprocessing step, we eliminate limit orders that never appear within ten levels of the best bid and ask prices.
In the experiment here, we use historical market data of GOOG during one trading day in August 2017. Our results for GOOG follow the same evaluation metrics as for synthetic data.

\textbf{Stock-GAN and baselines}: We show the performance of Stock-GAN and  compare it to VAE and DCGAN variant of our network. We show these results for price distribution (Figure~\ref{fig:goog_price}), quantity distribution (Figure~\ref{fig:goog_qnt}), and inter-arrival times (Figure~\ref{fig:goog_interarrival_1}, \ref{fig:goog_interarrival_2}---shown in two larger graphs for clarity). As earlier, we capture these differences quantitatively using the KS distance shown in Table~\ref{KSGOOG}.
\begin{table}[t]
\resizebox{.9\columnwidth}{!}{
\begin{tabular}{l c c c}  
\toprule
   & Real,S-GAN & Real,VAE & Real,DCGAN  \\
\midrule
Price    & 0.126 & 0.218 & 0.181         \\
Quantity  & 0.182  & 0.248 & 0.471     \\
Inter-arrival  & 0.066   & 0.835 & 0.154    \\
\bottomrule
\end{tabular}}
\caption{KS distances against real (GOOG)} \label{KSGOOG}
\end{table}
Similar to synthetic data, the numbers reveal that Stock-GAN is able to model GOOG data better than the baselines.
Intensity is measured in the same way as synthetic data, except we choose $T=1000$ seconds sized chunks of time due to the longer horizon of GOOG data. Figure~\ref{fig:goog_int} shows much smoother intensity produced by VAE and DCGAN as opposed to Stock-GAN which is much closer to the real data intensity.

\textbf{Ablation}: The real and Stock-GAN generated best bid/ask evolution are in Figures~\ref{fig:goog_bestba} and~\ref{fig:goog_bestbagen} respectively. As for synthetic data, we perform two ablation experiments, one by removing the CDA network (no cda) and one by removing order-book information (no ob), shown in Figure~\ref{fig:goog_bestbanocda} and~\ref{fig:goog_bestbanoob} respectively. 
The quantitative distinction is seen in the spectral densities for these time series shown in Figures~\ref{fig:goog_spectral}-\ref{fig:goog_spectralnoob}. 
However, unlike the synthetic data, here it can be seen that no cda has more higher frequency components that real data, which can also be seen by the high variation over time in Figure~\ref{fig:goog_bestbanocda}. On the other hand, no ob has less higher frequency (or even lower frequency) components which results in the flat shape in Figure~\ref{fig:goog_bestbanoob}. The Stock-GAN spectral density, while closest to real one among all alternatives, also misses out on some low frequency components. Nonetheless, Stock-GAN is closest to real data due to our novel structural approach of the CDA network and use of order-book data.

\section{Limitations and Conclusion}
We showed the superior performance of Stock-GAN in producing realistic market order streams compared to other approaches. In doing so, we also introduced five statistics to measure the realism of generated stock market order stream. We chose our real GOOG data for dates in which there were no external events, such as financial performance report. Thus, we did not model the effect of exogenous factors on stock market, which we believe is technically possible by just adding another condition for the generator. Notwithstanding these effects, we demonstrated that stock market data can be generated with high fidelity which provides a means for conducting research on sensitive stock market data without access to the real data.  In future work, we intend to test the effectiveness of the Stock-GAN on more stocks, other than PN and GOOG that we did in this work.

\subsection*{Acknowledgement}
We thank OneMarketData for the dataset used for this work. Most of this work was done when Junyi, Yaoyang, and Arunesh were at the University of Michigan, where the work was supported by US National Science Foundation under grant IIS-1741190.



\bibliography{ijcai19}
\bibliographystyle{aaai}
\end{document}